\documentclass[referee]{aa}

\usepackage{longtable}
\usepackage{graphicx}
\usepackage{txfonts}

\usepackage{natbib,twoopt}
\usepackage[breaklinks=false]{hyperref} %% to avoid \citeads line fills
\bibpunct{(}{)}{;}{a}{}{,}             %% natbib format like A&A and ApJ

\begin{document}
\title{Physical properties of B-type asteroids from WISE data}

\author
{
  V. Al\'i-Lagoa
  \inst{1}\fnmsep\inst{2}, 
  J. de Le\'on
  \inst{3}, 
  J. Licandro
  \inst{1}\fnmsep\inst{2}, 
  M. Delb\'o
  \inst{4}, 
  H. Campins
  \inst{5}, \\
  N. Pinilla-Alonso  
  \inst{6,7} 
  \and
  M. S. Kelley
  \inst{8}
}

\authorrunning
{
  V. Al\'i-Lagoa et al.
}

\institute
{
  Instituto de Astrof\'isica de Canarias (IAC), 
  c/ V\'ia L\'actea s/n, 38205, La Laguna, Tenerife, Spain
  \\
  \email{vali@iac.es}
  \and
  Departamento de Astrof\'isica, Universidad de La Laguna. 38206, La Laguna, Tenerife, Spain
  \and
  Departamento de Edafolog\'ia y Geolog\'ia,  Universidad de La Laguna. 38206, La Laguna, Tenerife, Spain
  \and
  UNS-CNRS-Observatoire de la C\^ote d'Azur, B.P. 4229, 06304 Nice Cedex 4, France
  \and
  Physics Department, University of Central Florida, P. O. Box 162385, Orlando, FL 32816.2385, USA
  \and
  Instituto de Astrof\'isica de Andaluc\'ia (IAA), Granada, Spain
  \and
  Department of Earth and Planetary Sciences, University of Tennessee, 1412 Circle Dr, Knoxville TN 37996-1410 
  \and
  Department of Astronomy, University of Maryland, College Park, MD 20472-2421, USA
}

\date{Received 31 October 2012; accepted 19 March 2013 (v1)}

\abstract
{}  % Context
{
  % Aims
  Our aim is to obtain more information about the physical nature of B-type asteroids and extend on the previous work by studying their physical properties derived from fitting an asteroid thermal model to their NASA's Wide-field Infrared Survey Explorer (WISE) data.   
  We also examine the Pallas collisional family, a B-type family with a moderately high albedo in contrast to the large majority of B-types. %, which have very low albedos.  
}
{
  We apply a combination of the Near-Earth Asteroid Thermal Model and a model of the reflected sunlight to WISE asteroid data in order to derive up to four parameters: effective diameter ($D$), the so-called infrared beaming parameter ($\eta$), ratio of infrared to visible albedo ($R_p = p_{IR}/p_V$) and visible geometric albedo ($p_V$). 
}
{
  We obtained the effective diameter, geometric visible albedo, infrared-to-visible albedo ratio and beaming parameter for $\ga$ 100 B-types asteroids and plotted the value distributions of $p_V$, $R_p$ and $\eta$ ($\bar{p}_V = 0.07 \pm 0.03$, $\bar{R}_p = 1.0 \pm 0.2$, and $\bar\eta = 1.0 \pm 0.1$). 
By combining the IR and visible albedos with 2.5 $\mu$m reflectances from the literature we obtained the ratio of reflectances at 3.4 and 2.5 $\mu$m, from which we found statistically significant indications that the presence of a 3-$\mu$m absorption band related to water may be commonplace among the B-types. 
  Finally, the Pallas collisional family members studied ($\sim$ 50 objects) present moderately high values of $p_V$ ($\bar{p}_V = $ 0.14 $\pm$ 0.05), significantly higher than the average albedo of B-types. 
  In addition, this family presents the lowest and most homogeneously distributed $R_p$-values of our whole sample, which shows that this group is clearly different from the rest of B-types, likely because its members are pieces likely originating from the same region of (2) Pallas, a particularly high-albedo B-type asteroid. 
}
{}


\maketitle
% 
% ________________________________________________________________

\section{Introduction \label{sec:intro}}

The study of asteroids is important to gain knowledge about the origin and evolution of our planetary system. 
Asteroids are relics of the Solar System's formation and the building blocks of the terrestrial planets.
Primitive asteroids, i.e. those belonging to the so-called spectroscopic C-complex and having in general visible geometric albedo $p_V \la 0.1$ and featureless, flat visible spectra, are particularly relevant in this context. 
Thought to have formed further away from the sun than the other asteroid classes, primitive asteroids have experienced less heating and alteration processes 
and have a more pristine composition, potentially preserving crucial information about the early Solar System. 
In addition, primitive asteroids play an important role in current exobiological scenarios as they delivered complex organic molecules to the early Earth.
This organic matter is prerequisite for the synthesis of pre-biotic biochemical compounds that would subsequently lead to the emergence of life \cite[][and references therein]{Maurette2006}.
For these and other reasons, upcoming sample return space missions have selected primitive asteroids as primary targets: NASA's OSIRIS-Rex \citep{Campins2010,Lauretta2010}, ESA's Marco Polo-R \citep{Barucci2012,deLeon2011}; and JAXA's Hayabusa-2\footnote{\tt http://www.jspec.jaxa.jp/e/activity/hayabusa2.html}.

The taxonomic classification of primitive asteroids has been traditionally based on their low visible albedo ($\la$\,0.08--0.1), relatively flat or slightly blue visible spectra and the weak or no absorption features thereof \citep[for a detailed review, see][]{Clark2010}.
Several primitive classes were defined in Tholen's taxonomy, e.g. B, C, F, G, D and P \citep{Tholen1984,Tholen1989}. 
Bus' feature-based classification, independent of the albedo, merged some of these and defined new primitive taxons that were extended with minor changes into the Bus-DeMeo taxonomy \citep{Bus2002a,DeMeo2009}. 

B-types are of particular interest among the primitive asteroids for a number of reasons: 
(i) there is as yet no compelling explanation for their defining feature, i.e. their slightly blue spectral slope in the visible range; 
(ii) B-type asteroids constitute the only primitive class that presents a wide range of spectral slopes in the 0.8--2.5 $\mu$m near-infrared region, from negative to positive, in the 0.8--2.5 $\mu$m near-infrared region \citep{deLeon2012}; 
(iii) the few B-types studied present the 3-$\mu$m absorption feature related to hydrated minerals; 
(iv) water ice has been detected on the surface of (24) Themis \citep{Campins2010a,Rivkin2010}; 
(v) the majority of asteroids that have been observed to display cometary-like activity are B-types \citep[][and references therein]{Licandro2012}; 
(vi) the target of NASA OSIRIS-Rex mission, 2006\,RQ$_{36}$, is a B-type asteroid. 

B-type asteroids have been widely related to carbonaceous chondrites, composed of carbonaceous minerals and phyllosilicates, in terms of their generally low albedo and broad spectral properties \citep{Gaffey1989,Vilas1989,Vilas1994}. 
More recently, \citet{deLeon2012} examined visible to near-infrared (VNIR) spectra of a sample of 45 B-types and found that the characteristic negative spectral slope in visible wavelengths diverges into a continuum of gradually varying spectral slopes in the 0.8--2.5 $\mu$m (NIR) range, from a monotonic negative (blue) slope to a positive (red) slope. 
\citet{deLeon2012} classified their spectra into six ``average spectra'' or ``centroids'' representative of the whole sample by means of statistical clustering analysis \citep{Marzo2009}.
These centroids were compared against meteorite spectra from the RELAB database \citep{Pieters2004}. 
The best meteorite analogues found for the six ``clusters'' were all carbonaceous chondrites with a gradual change in their degree of hydration, from aqueously altered CM2 chondrites for the reddest cluster, to the heated/thermally metamorphosed CK4 chondrites for the bluest one.

This work is an extension of the results obtained by \citet{deLeon2012}, which are part of a ongoing programme devoted to improving our knowledge of B-types. 
Our aim here is to study the physical properties of B-type asteroids that can be derived by fitting a thermal model to their NASA's Wide-field Infrared Explorer (WISE) observations, i.e. effective diameter, beaming parameter and $p_{IR}/p_V$, where $p_{IR}$ is the albedo at 3.4--4.6 $\mu$m as defined in \citet{Mainzer2011b}. 
We closely follow the methodology of \citet{Mainzer2011b}, though with a number of differences, as described in Sects. \ref{sec:data} and \ref{sec:model}. 

\citet{Mainzer2011e} studied all groups of spectrophotometrically classified asteroids in the Tholen, Bus and Bus-DeMeo taxonomies observed by WISE, including B-types. 
Those authors present visible and NIR geometric albedo distributions and median values of B-types and conclude that, in spite of having analogously low albedos, B-, C-, D- and T-type asteroids can be discriminated from their values of NIR reflectance. 
In particular, \citet{Mainzer2011e} point out that B-types have a lower $p_{IR}/p_V$ ratio than C-types and attribute this to their blue VNIR slopes likely extending out to 3--4 $\mu$m. 
Our definition of B-type asteroid in this work is different: following \citet{Clark2010} and \citet{deLeon2012}, we consider all objects that have a flat to slightly blue spectral slope in the visible range, i.e. any object that has ever been classified as a B-type, including Tholen's F-types and ambiguous designations. 
This criterion produces a total of 162 asteroids classified as B-types. 

We also study the collisional family of (2) Pallas (hereafter PCF and Pallas, respectively). 
The PCF is interesting for the following reasons: 
(i) it is a B-type family, given that Pallas and the very few family members that have been taxonomically classified are B-types \citep{Gil-Hutton2006,deLeon2010a}, and the five members studied in \citet{deLeon2012} were spectrally related to carbonaceous chondrites, which establishes their primitive nature;  
(ii) the average values of geometric albedo of members of the family calculated by \citet{Masiero2011} are roughly 0.15 (see their Fig. 19), significantly greater than expected for primitive bodies ($<0.1$), though no explicit comment is made by these authors on this intriguing result; 
(iii) the Near-Earth asteroid (3200) Phaethon, an activated asteroid parent of the Geminid meteor shower, likely originated in the Pallas family \citep{deLeon2010a};
(iv) this family is well isolated in (proper) element space, thus the potential identification of interlopers as members is greatly reduced. 

The paper is organised as follows. 
In Sect. \ref{sec:data}, we briefly describe the WISE data set and our selection criteria. 
The thermal modelling of the data is explained in detail in Appendix \ref{app:model}, whereas Sect. \ref{sec:model} includes relevant comments on the very few differences introduced in this work. 
In Appendix \ref{app:paramcomp} our parameter determinations are compared to those by \citet{Masiero2011}. 
We present our results in Sect. \ref{sec:results}, a discussion of the implications of this work is put forward in Sect. \ref{sec:discussion}, and our conclusions are enumerated in Sect. \ref{sec:conclusions}.

\section{Data \label{sec:data}}

A general introduction to WISE can be found in \citet{Wright2010} and references therein. 
Of particular interest to Solar System science is the NEOWISE project. 
This acronym collectively refers to two enhancements to the WISE data processing system that were designed to allow detection and archiving of Solar System objects \citep[for details, see][]{Mainzer2011a}.    

WISE used four broad-band filters with approximate isophotal wavelengths at 3.4, 4.6, 12 and 22 $\mu$m, referred to as W1, W2, W3 and W4, respectively \citep{Wright2010}.
The WISE All-Sky Single Exposure L1b Working Database, published in April 2012 and available via the IRSA/IPAC archive\footnote{http://irsa.ipac.caltech.edu/Missions/wise.html}, includes the corresponding magnitudes and uncertainties in the Vega system as well as quality and contamination and confusion flags that enable us to reject defective data \citep{Cutri2012}.  

We follow a combination of criteria found in \citet{Mainzer2011b,Mainzer2011e,Masiero2011} and \citet{Grav2012} in  order to ensure the reliability of the data. 
We implement the correction to the red and blue calibrator discrepancy in W3 and W4; 
we use a cone search radius of 0.3\arcsec\, centred on the MPC ephemeris of the object in our queries; 
all artifact flags other than p, P and 0 and quality flags other than A, B and C are rejected; 
we require the modified Julian date to be within 4 seconds of the time specified by the MPC and split groups of epochs separated more than three days (see the end of this section);  
we ensure that the data is not contaminated by inertial sources by removing those points that return a positive match from the WISE Source Catalog within 6\arcsec; 
finally, all remaining observations in a given band are rejected if they are fewer than 40\% of the data in the band with the maximum number of detections. 

On the other hand, we do not use data saturated to any extent. 
The onset of saturation is reported to correspond to magnitudes $M_{\mathrm{W1}} < 6$, $M_{\mathrm{W2}} < 6$, $M_{\mathrm{W3}} < 4$, $M_{\mathrm{W4}} < 3$ \citep{Cutri2012}. 
We found that enlarging the error bar of partially saturated data to 0.2 magnitudes (which translates into a relative error of 20\% in fluxes) renders the corresponding band to play no effective role in the thermal model fit by not contributing significantly to the $\chi^2$. 

The application of the above criteria results in a sample of 111 B-type Main-Belt objects with WISE observations usable for our purposes.
Some asteroids have been observed by WISE in more than one uninterrupted group of epochs with different observation geometries. 
We also model separately such groups of observations if they are more than three days apart \citep[see Appendix \ref{app:model} and ][]{Mainzer2011b}. 
Consequently, we have a larger set of parameter determinations than asteroids in the sample.

\section{Thermal modelling \label{sec:model}}

The modelling of WISE asteroid data implemented in this work closely follows \citet{Mainzer2011b, Mainzer2011e, Masiero2011} and is based on the Near-Earth Asteroid Thermal model \citep[NEATM, ][]{Harris1998} and the IAU phase curve correction to the visible magnitude \citep{Bowell1989}. 
 For the sake of reproducibility, we include a detailed account of our procedure in Appendix \ref{app:model} and enumerate the few differences with respect to Masiero et al. and Mainzer et al. below. 

The number of parameters we can fit for each object depends upon how many and which WISE bands are present in its data set.
Parameter default values are chosen based on the peak of their respective fitted value distributions of Main Belt Asteroids presented in \citet{Masiero2011}. 
Whenever there is one single or no thermal band available (W2, W3 or W4) we assume $\eta=1$.0; $R_p$ is is fixed to 1.5 unless we have at least 50\% contribution of reflected sunlight in W1 data. 
The last criterion is based on the consistency of our parameter determinations for objects with double detections\footnote{This criterion may not be of general applicability and has only been checked for objects in our B-type sample.}. 
Namely, both groups of observations of asteroids (1076) and (2446) have $> 50$\% sunlight and their $R_p$ values are consistent within the errorbar. 
On the other hand, (3579) has non-compatible $R_p$ determination from W1 data with  $>$70\% and $\sim$ 25\% reflected sunlight, respectively. 
We thus reject three $R_p$ values belonging to asteroids (288), (1493) and (3579). 
 
These considerations allow asteroid size to be fitted in all cases and, by means of the relation
\begin{equation}
  p_V = \left(1329\,\left[\mathrm{km}\right] \frac{10^{-H/5}}{D\,\left[\mathrm{km}\right]}\right)^2,
  \label{ec:pVHD}
\end{equation}
the geometric visible albedo can be computed. 
In contrast, W1 and W2 data are more often rejected based on the data requirements (see Sect. \ref{sec:data}) than the purely thermal bands and one will usually be able to obtain fewer $R_p$ determinations than $\eta$ or indeed $D$. 

\begin{figure}
  \centering
  \includegraphics[width=68mm]{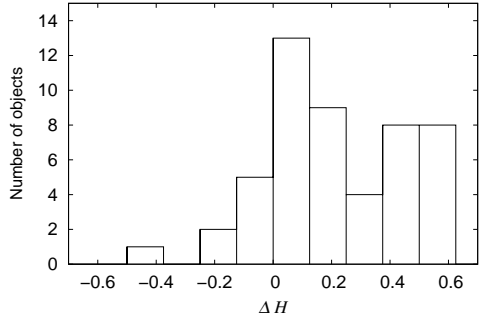}
  \caption{
    Differences in absolute magnitude values between those used by  \citet{Masiero2011}, $H_M$, and the most updated ones (as of May 2012) used in this paper, $H_U$. 
    Note that the cases verifying $\Delta H = 0$ do not contribute to this histogram.
  }
  \label{fig:HmagDiff}
\end{figure}

It is important to point out that we do not use physical data previously determined by direct measurements such as radar diameters or albedos to constrain our fits, i.e. we limit ourselves to using radiometrically derived sizes. 
This will introduce variations on the parameter determinations of some individual asteroids --specially the largest ones since it is more likely that more direct measurements have been performed-- as compared to \citet{Masiero2011}, but should not affect the result of statistical analyses if the populations studied consist of a significant number of objects. 
In Appendix \ref{app:paramcomp} we show that, given the same input values of $H$, our best-fit parameter values are consistent within the errorbars, though we find that our $R_p$ determinations are systematically lower by $\sim$10\%. 
On the other hand, the update of $\sim 45$\% of the MPC $H$-values in our sample does change the values of $p_V$ (see Eq. \ref{ec:pVHD}) and $R_p$. 
As shown in Fig. \ref{fig:HmagDiff}, the updates tend to be toward greater values of $H$, which will result in lower $p_V$ and greater values of $R_p$ than those of \citet{Masiero2011}. 
For more details, see Appendix \ref{app:paramcomp}. 

\section{Results \label{sec:results}}

\subsection{Value distributions of $\eta$, $p_V$ and $R_p$ \label{sec:parametercomp}}

\begin{figure*}
  \begin{center}
    \includegraphics[width=68mm]{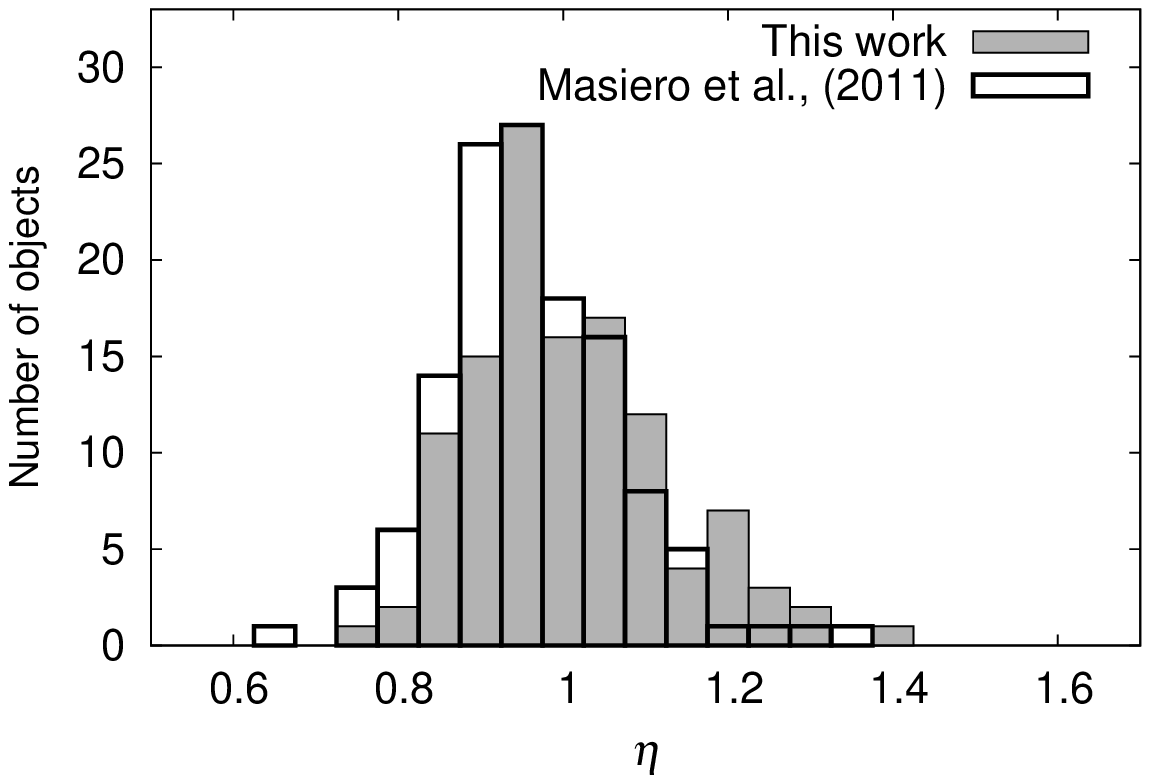}
    \includegraphics[width=68mm]{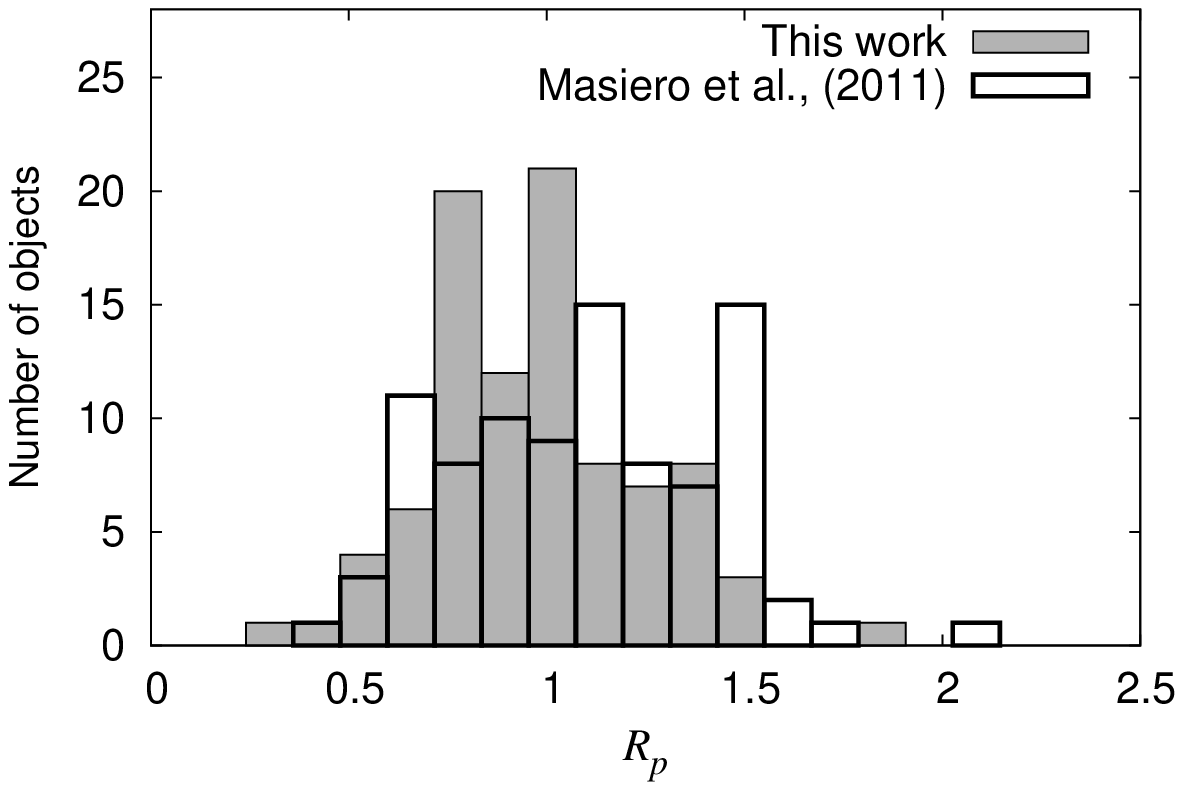}

    \includegraphics[width=68mm]{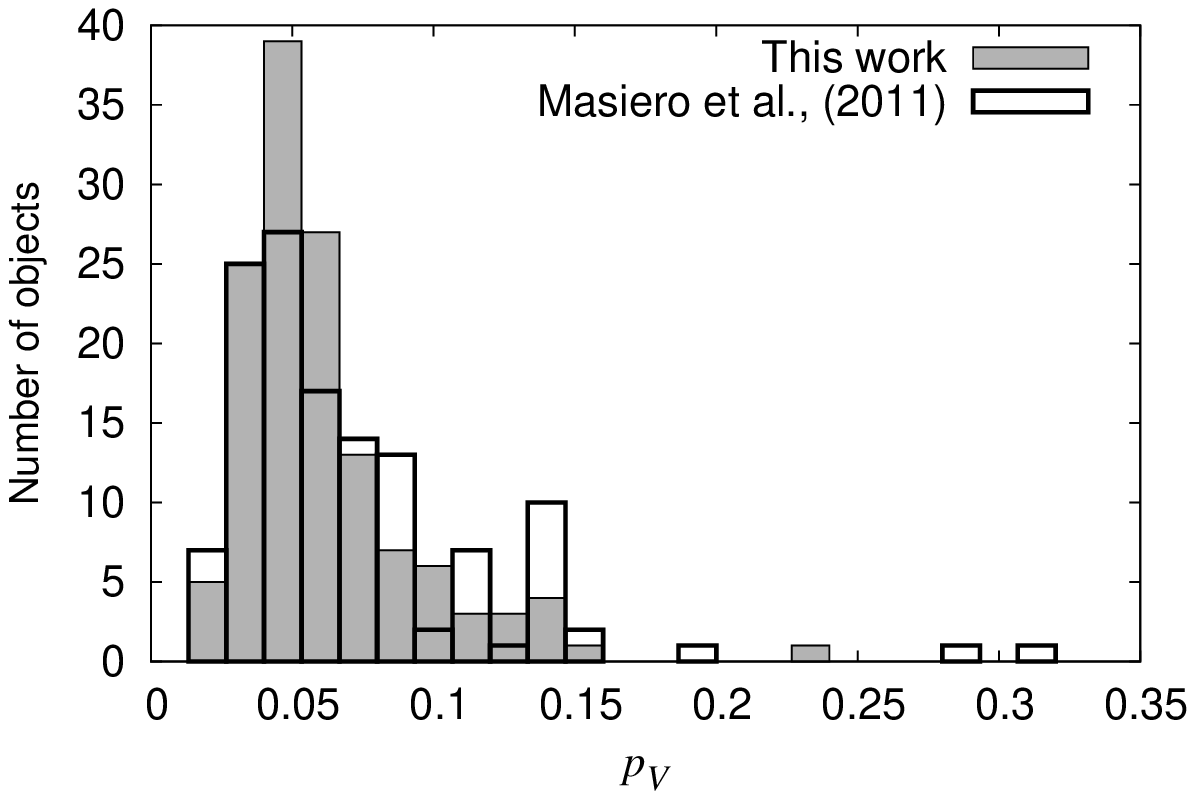}
    \includegraphics[width=68mm]{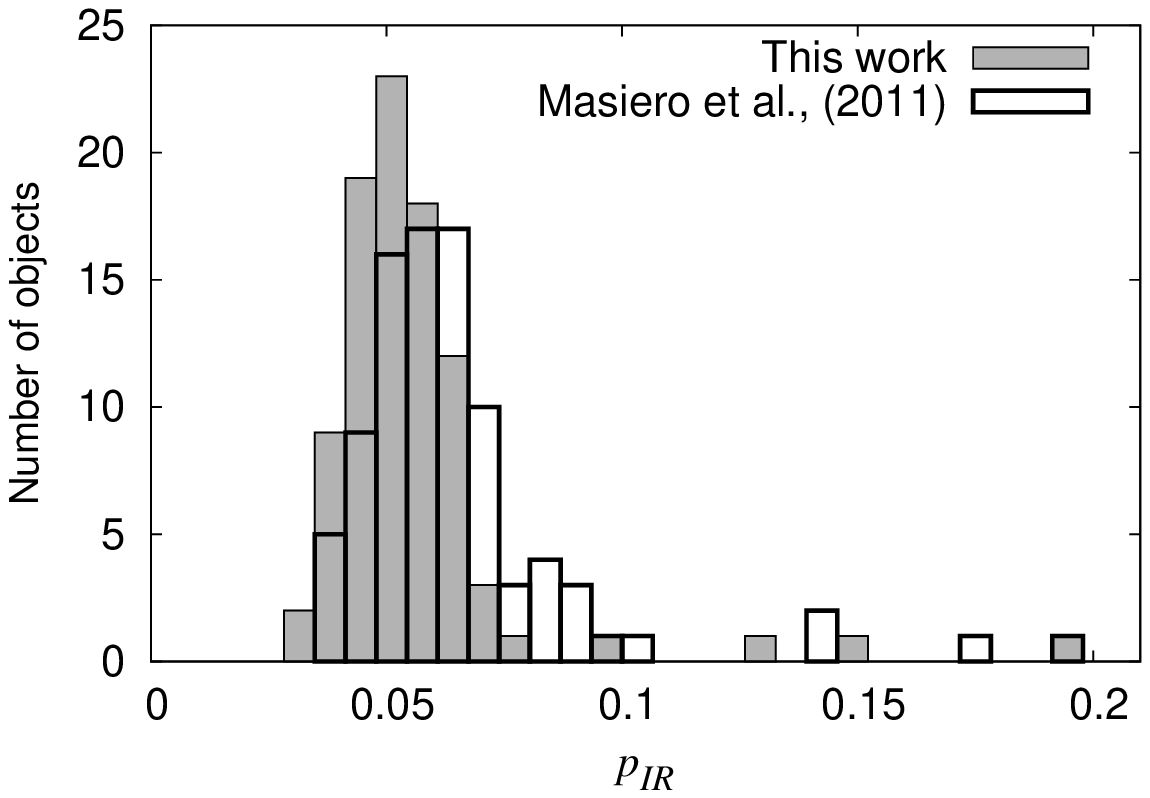}
  \end{center}
  \caption{
    Histograms of value distributions obtained for the B-type asteroids observed by WISE: $\eta$ (top left), $p_V$ (bottom left), $R_p$ (top right), and $p_{IR}$ (bottom right). 
    The corresponding histograms using the results from \citet{Masiero2011} are overplotted. 
  }
  \label{fig:histograms}
\end{figure*}

The distributions of $\eta$, $p_V$, $R_p$ and $p_{IR}$ obtained for the B-types are shown in Fig. \ref{fig:histograms} (the complete set of parameters determinations is given in Table \ref{supertable}\addtocounter{table}{1}). 
For comparison, we overplotted the corresponding histograms with best-fit parameter values from \citet{Masiero2011} (see Appendix \ref{app:paramcomp} for a detailed comparison). 
The parameter median and mean values and standard deviations of this work, as well as the number of parameter determinations obtained in each case ($N$), are presented in Table \ref{table:mean}. 
Note that $p_V$ and $p_{IR}$ are not fitted, but computed. 
The former is obtained from Eq. \ref{ec:pVHD} with the best-fit value of $D$ as input, whereas $p_{IR} = R_p p_V$.
These results are consistent with previous work by \citet{Mainzer2011e}: if we take the weighted mean of median $p_V$- and $R_p$-values corresponding to their Tholen, Bus and Bus-DeMeo B-types and Tholen F-types we obtain the same median values. 

We find a $\eta$-value distribution centered at unity, consistent with the average value obtained for the whole Main Belt \citep{Masiero2011}. 
The broad and asymmetrical $p_V$ distribution extends to $p_V > 0.1$. 
The $R_p$ distribution is also broad, whereas the values of $p_{IR}$ are more compactly distributed around the mean. 
\begin{table}[h]
  \caption{Median and mean values and standard deviations of $\eta$, $R_p$, $p_V$ and $p_{IR}$ derived for the B-type asteroids observed by WISE.} 
  \label{table:mean}      
  \centering                          
  \begin{tabular}{c c c c c}   
    \hline\hline                 
    Parameter & Median & Mean & $\sigma$ & $N$ \\    
    \hline                               
    $\eta$ & 1.0  &  1.0  & 0.1  & 116 \\      
    $R_p$  & 1.0  &  1.0  & 0.2  & \,88  \\
    $p_V$  & 0.06 &  0.07 & 0.03 & 132 \\
    $p_{IR}$  & 0.06 &  0.06 & 0.01 & \,88 \\
    \hline                                   
  \end{tabular}  
\end{table}

\subsection{Albedo ratio}
\label{sec:resultsRp}

\begin{figure}[h] 
  \centering
  \includegraphics[width=68mm]{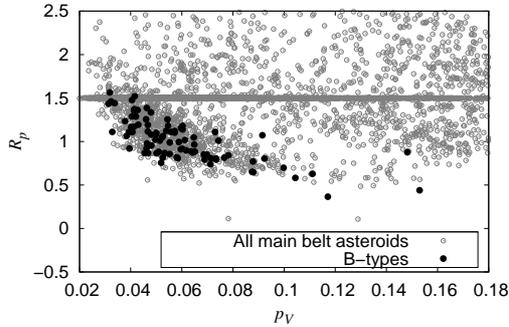}
  \caption{
    Albedo ratio versus visible geometric albedo. The 20\% errorbars in $p_V$ and $R_p$ are not shown to facilitate visualisation. 
    B-types observed by WISE are plotted in black circles;  all main belt objects featured in Table 1 of \citet{Masiero2011} are in empty grey circles.  
    The horizontal line is an artificial feature corresponding to objects with the default fixed value of $R_p$=1.5. 
  }
  \label{fig:rpvspvAll}
\end{figure}

Figure \ref{fig:rpvspvAll} shows a plot of $R_p$ versus $p_V$. 
Similar plots including all taxonomic classes in different classification schemes are presented in Figs. 14 and 15 by \citet{Mainzer2011e} to show how clearly different taxons may be distinguished. 
Here we concentrate on the $p_V < 0.18$ range, with all B-types with WISE data for which $R_p$-values could be derived are plotted in black circles (see Table \ref{table:mean}); we also include all main belt asteroids taken from Table 1 of \citet{Masiero2011} (grey empty circles). 
The cloud of points exhibits a characteristic  ``waning-moon'' shape, with no points in either the high-$p_V$, high-$R_p$ or low-$p_V$, low-$R_p$ regions of the plot. 
Mainzer et al. caution that, while WISE is essentially unbiased against $p_V$, spectroscopic surveys conducted to create the classification schemes are inherently biased against small, low-$p_V$ objects; in addition, the computation of $R_p$ from WISE data requires sufficient reflected sunlight contribution in bands W1 and W2, which will tend to exclude objects with low enough values of $p_V$ and $R_p$. 
This could explain the lack of points in the lower left part of the plot. 
However, these biases cannot be solely responsible for this characteristic shape since other taxonomic classes with higher values of $p_V$ and $R_p$ also cluster similarly. 
Furthermore, if we plot all main belt objects irrespective of whether they have a taxonomic classification or not, distinct clouds of points with the same shape become apparent. 
Thus, we emphasize that because of their characteristic $p_V$- and $p_{IR}$-value distributions, there are no high-$R_p$ objects among the high-$p_V$ B-type asteroids. 

\begin{figure}
  \centerline{\includegraphics[width=68mm]{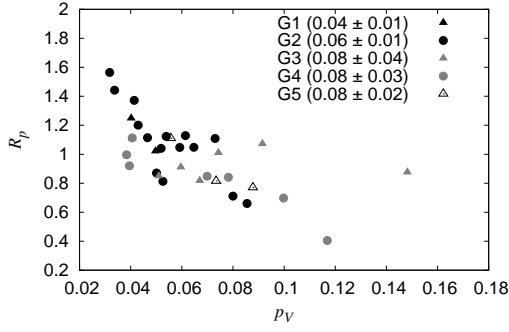}}
  \caption{
    $R_p$ vs. $p_V$ for the asteroids in \citet{deLeon2012} distinguishing the clusters to which they belong. 
    Within parentheses, the average $p_V$ of each cluster is shown. 
    On average, $p_V$ increases from G1 to G3, keeping the same value from G3 to G5.
  }
  \label{fig:Rpclusters}
\end{figure}

We have also analysed the $R_p$-values of the sample of 45 B-types studied by \citet{deLeon2012} separately. 
The spectra of these asteroids were classified into six ``average spectra'' or ``centroids'' referred to as G1, G2, ... G6 (see Sect. \ref{sec:intro}). 
These show a progressive decrease in spectral gradient in the NIR interval (0.8--2.5 $\mu$m), ranging from a positive (red) slope for G1 to a negative (blue) slope for G6.  
In Fig. \ref{fig:Rpclusters}, we plot $R_p$ versus $p_V$ labelling the objects in the different centroids G1, $\ldots$ G5. 
Note that cluster G6 is not included since it is composed only of one member, (3200) Phaethon, which did not have enough WISE observations to perform a reliable fit.
The average values of $p_V$ for each cluster increases from G1 to G3, retaining the same value from G3 to G5. 
This might suggest an inverse correlation between the cluster NIR slope and $p_V$, though the small number of objects per cluster with WISE observations (2 objects in G1, 12 in G2, 6 in G3, 5 in G4, and 3 in G5) prevents us from stablishing a firm conclusion. 

\begin{figure}
  \centerline{\includegraphics[width=68mm]{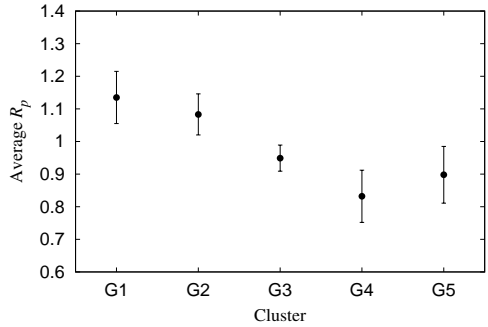}}
  \caption{
    Average values of $R_p$ for each cluster defined by \citet{deLeon2012}. In terms of spectral slope up to 2.5 $\mu$m, cluster G5 is the bluest, whereas G1 is the reddest. 
    Note: the errorbar is the standard error of the mean. 
  }
  \label{fig:averRpclusters}
\end{figure}

In Fig. \ref{fig:averRpclusters}, the average $R_p$-value is plotted for the different clusters. 
This figure suggests a correlation between the average $R_p$ and the NIR slope of the relative reflectance of the clusters: objects with higher $R_p$ belong (on average) to clusters with higher NIR spectral slope in the $\sim$ 1.0--2.5 $\mu$m range \citep[see also Fig. 5 of ][]{deLeon2012}. 
This indicates that the reflectivity at 3.4 $\mu$m tends to continue the trend observed at shorter IR wavelengths, as hypothesized by \citet{Mainzer2011e}.  

The W1 band pass spans the 2.8 to 3.8 $\mu$m range \citep{Wright2010}; therefore, the values of $R_p$ may also be diagnostic of the presence of the 3-$\mu$m absorption feature attributed to hydrated minerals or water-ice detected on many asteroids \citep{Rivkin2000,Gaffey2002,Campins2010,Rivkin2010,Licandro2011}. 
\citet{Mainzer2011e} ruled out the possibility of detecting the hydration band from WISE data based on the fact that the average $R_p$ for a sample of 7 M-types with positive detections of the band \citep{Rivkin2000} cannot be distinguished from that corresponding to other 33 M-types.  
However, this test might not be meaningful given that the $R_p$-values have errorbars at least larger than the characteristic depth of the absorption feature.  
Below we provide evidence that the W1 may be sensitive to the 3-$\mu$m feature. 

\begin{figure*}
  \begin{center}
    \includegraphics[width=0.32\linewidth]{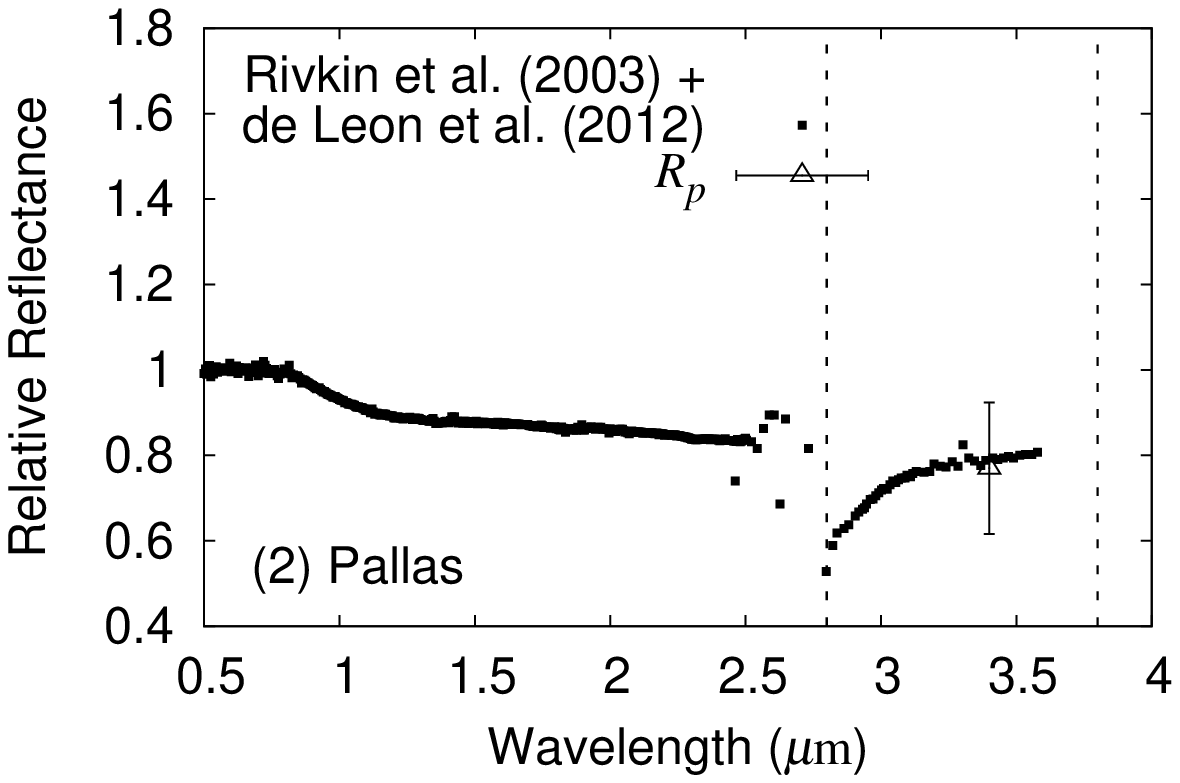}
    \includegraphics[width=0.32\linewidth]{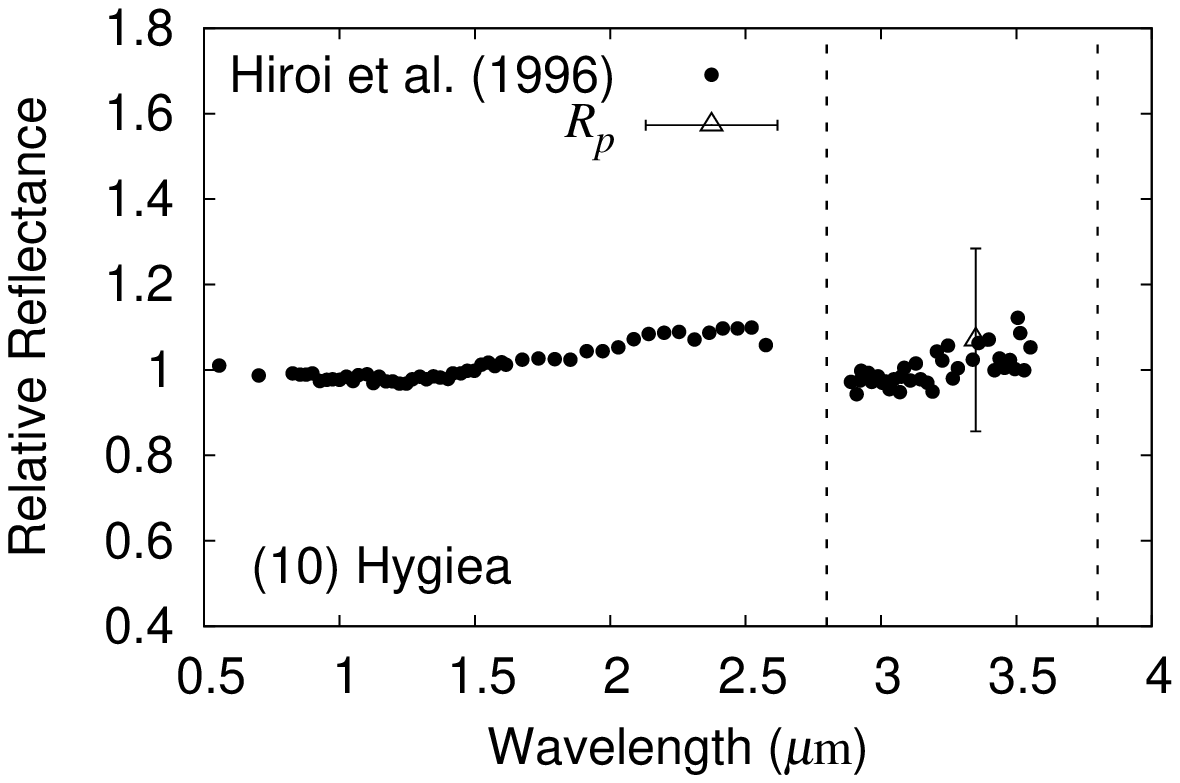}
    \includegraphics[width=0.32\linewidth]{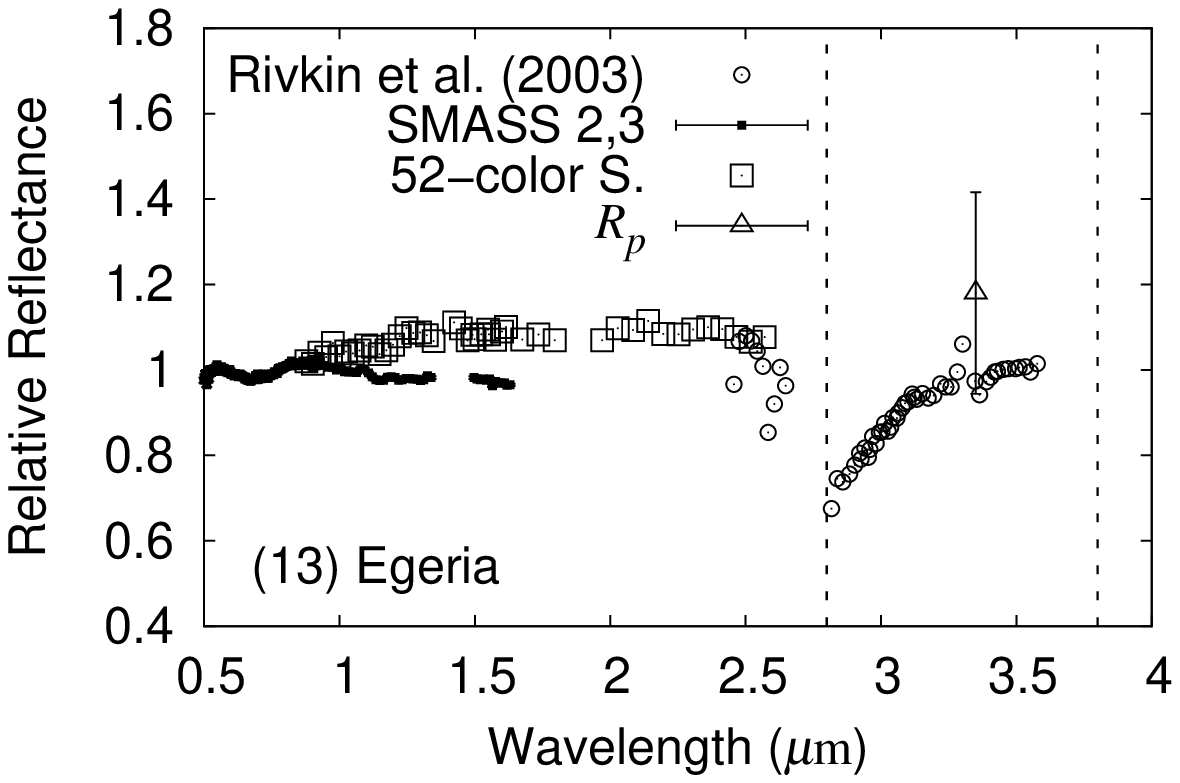}
    
    \vspace{-0.3cm}
    \includegraphics[width=0.32\linewidth]{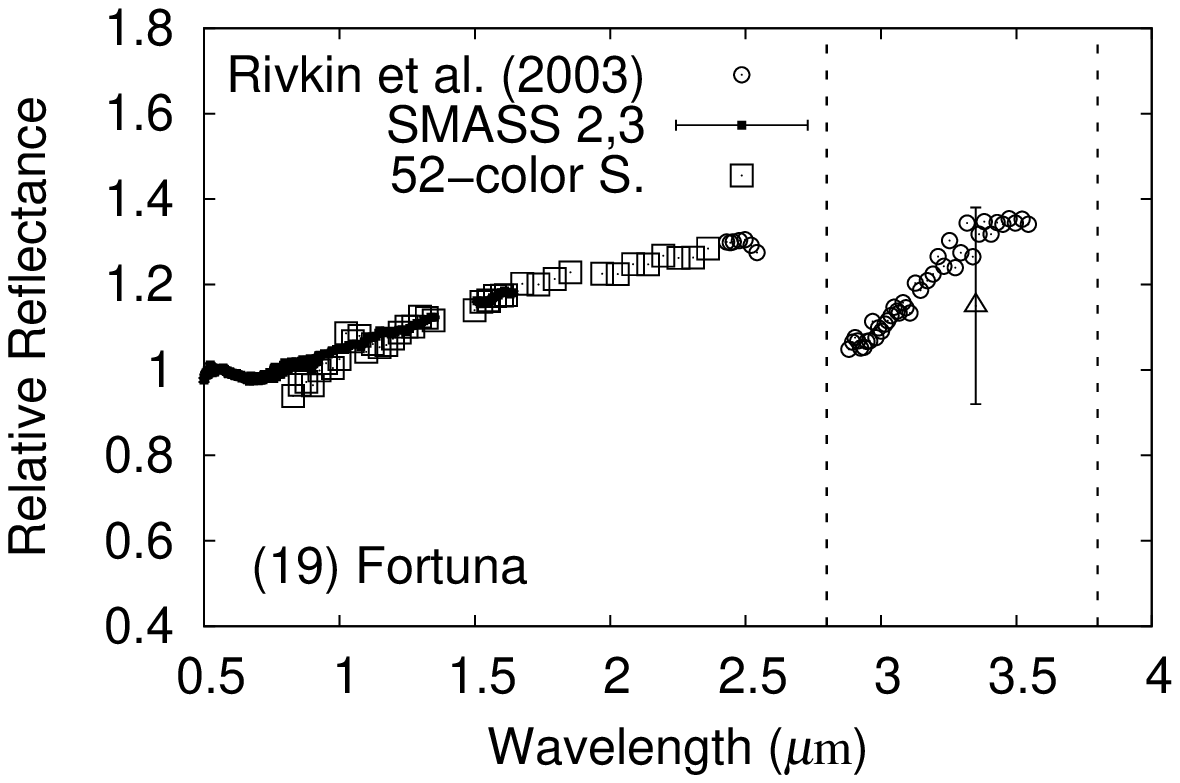}
    \includegraphics[width=0.32\linewidth]{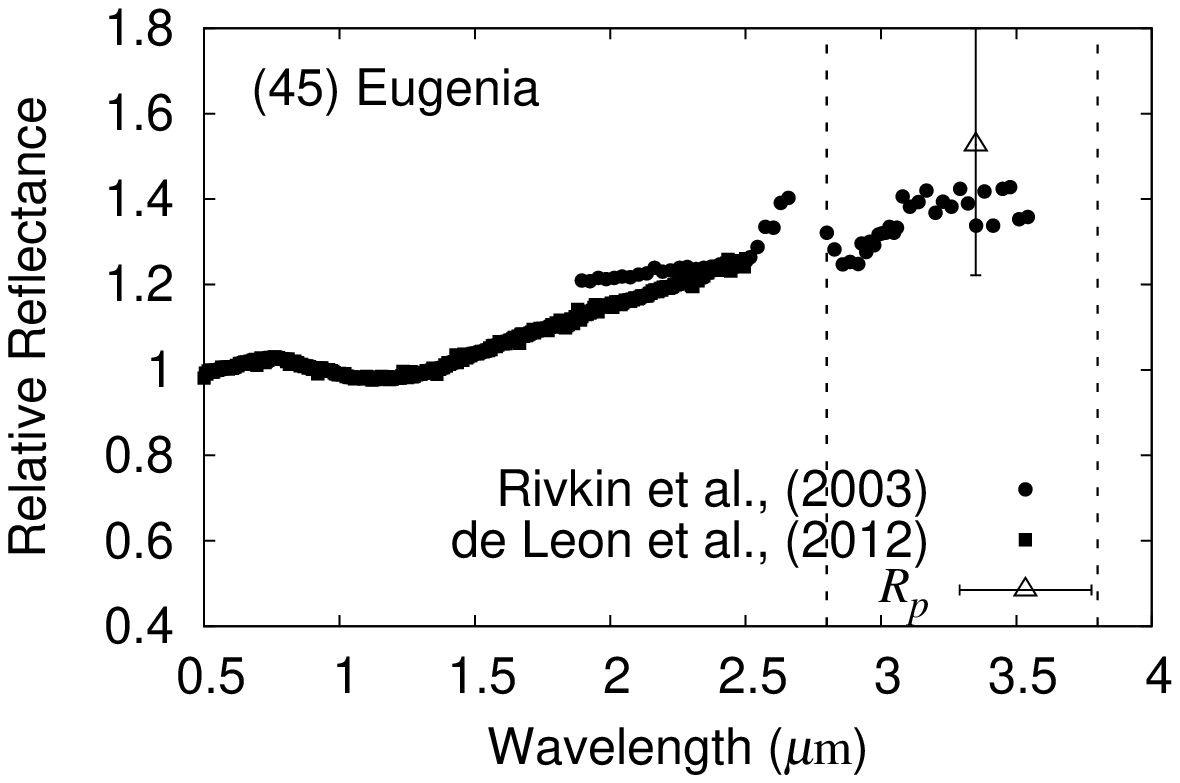}
    \includegraphics[width=0.32\linewidth]{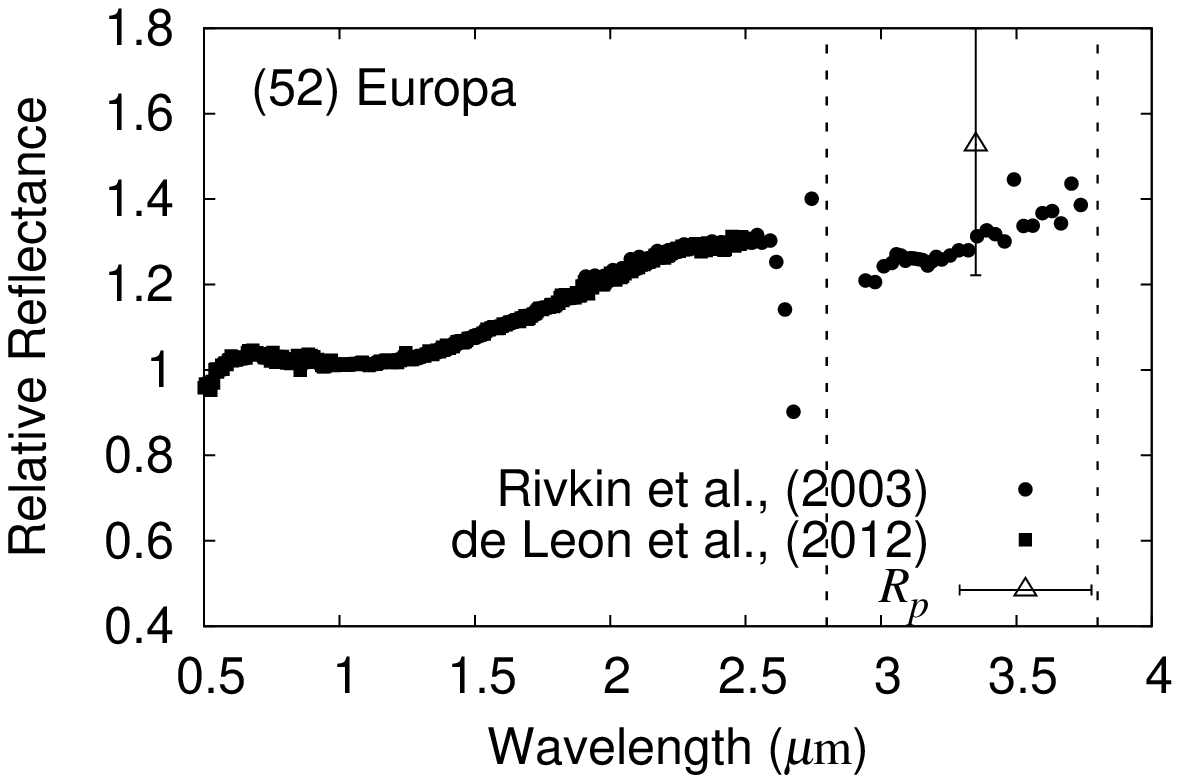}
 
    \vspace{-0.3cm}
    \includegraphics[width=0.32\linewidth]{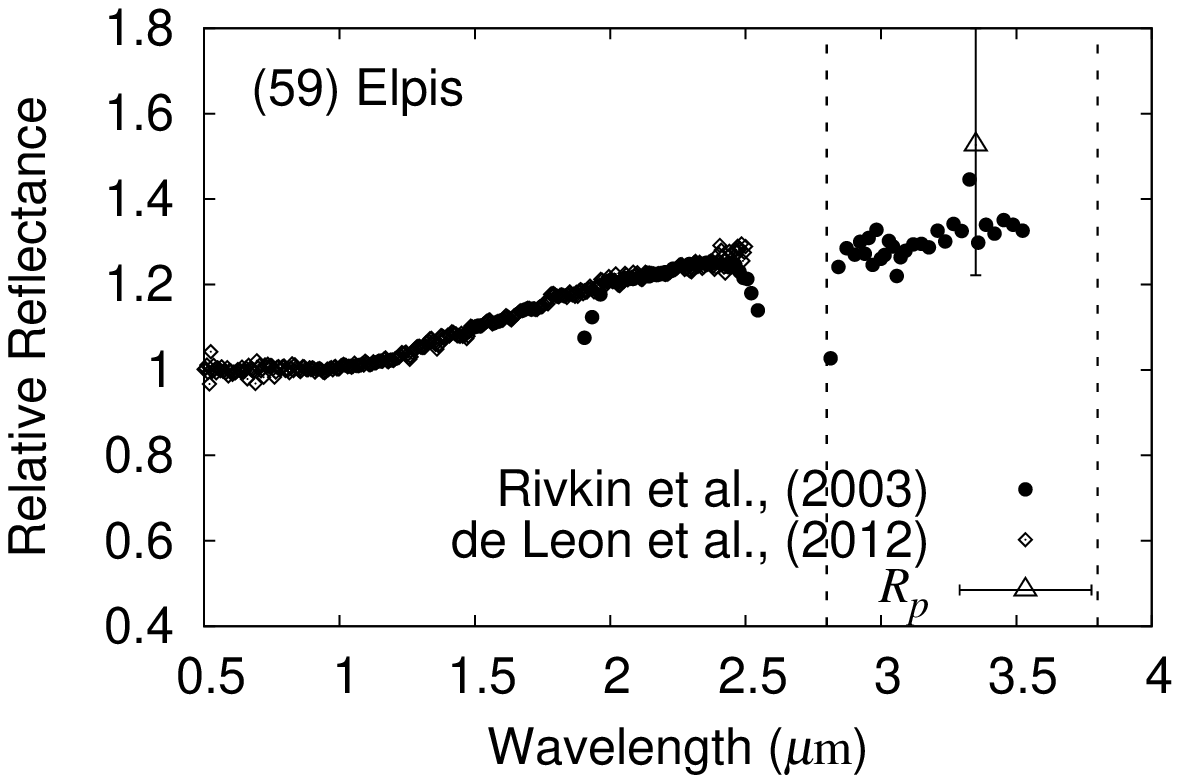}
    \includegraphics[width=0.32\linewidth]{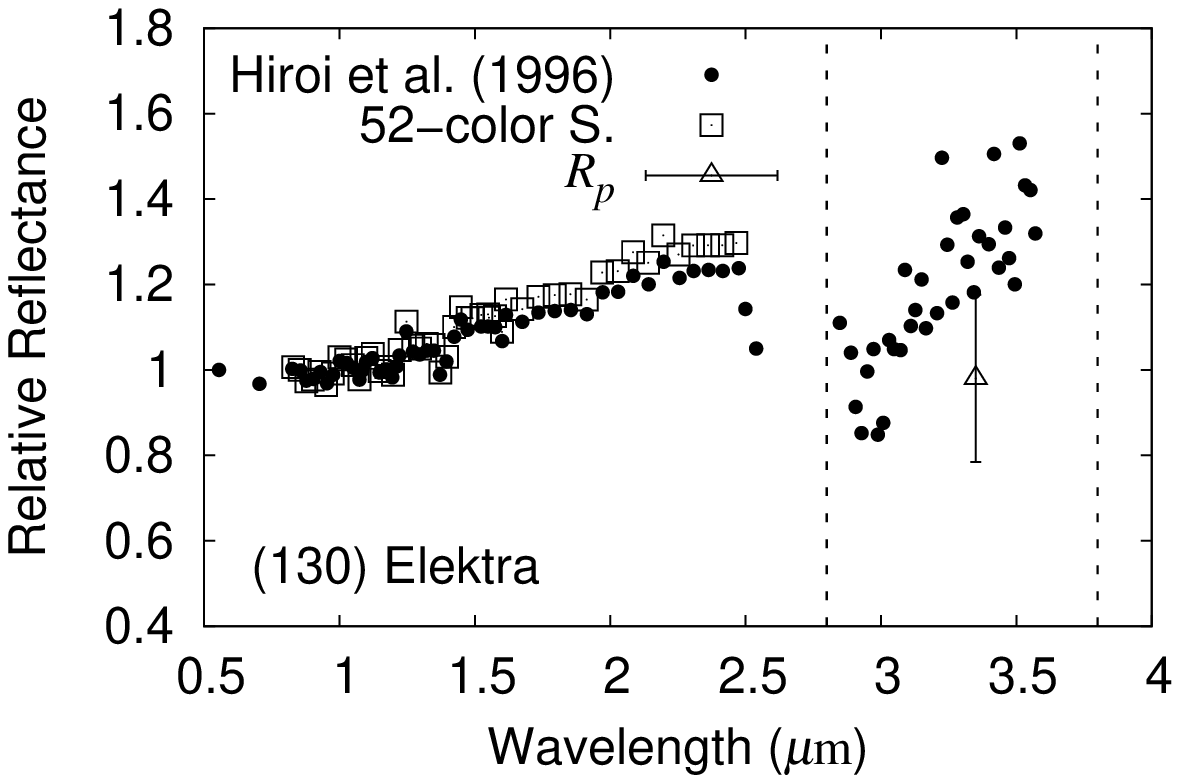}  
    \includegraphics[width=0.32\linewidth]{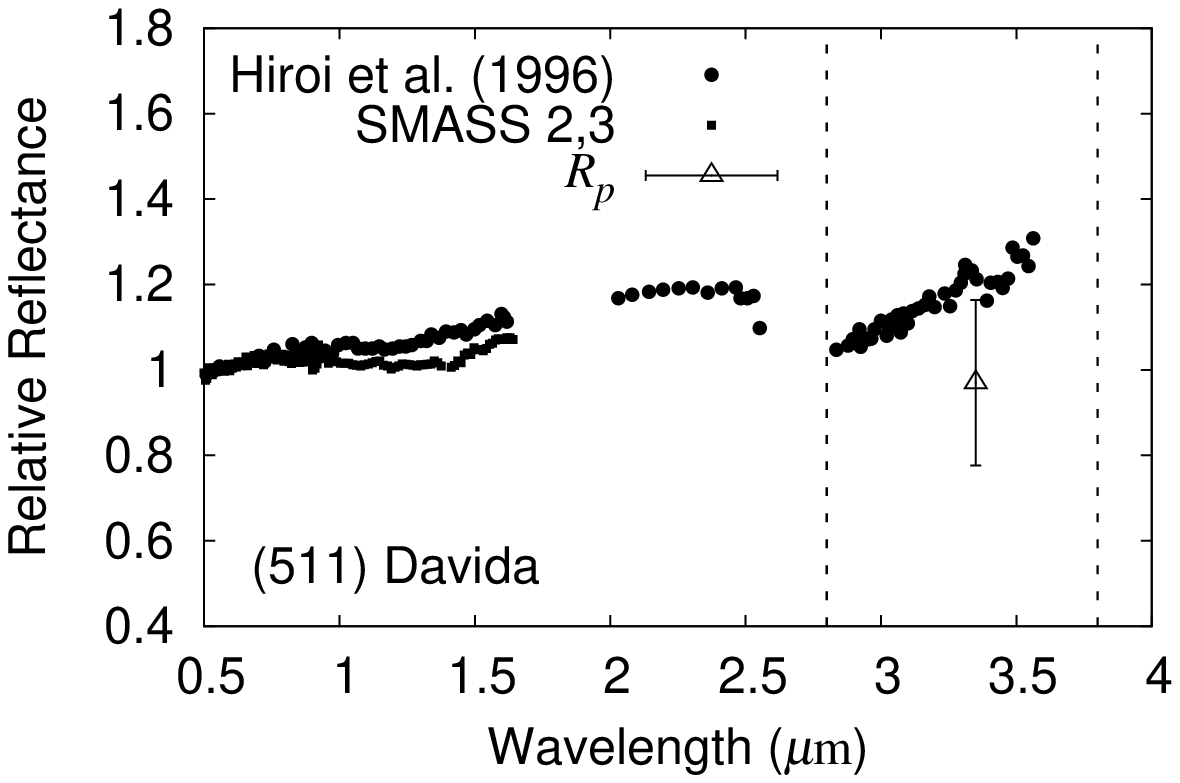}
  \end{center}
  \caption{
    Assembled spectra of the control objects chosen to study the reflectivity at 3.4 $\mu$m derived from WISE data as diagnostic of the presence of the 3-$\mu$m feature. 
    The dashed lines approximately enclose the wavelength integration range of the W1 filter. 
  }
  \label{fig:assembled}
\end{figure*}

Our first step was to assemble VNIR spectra up to $\sim 3.6\,\mu$m of a list of nine C-complex control asteroids. 
All of these asteroids are primitive, including some B-types, such as (2) Pallas or (45) Eugenia, and some of them show a distinct absorption feature. 
The assembled spectra are plotted along with the best-fit value of $R_p$ in Fig. \ref{fig:assembled}.
We collected or digitised data from \citet{Hiroi1996}, SMASS-II \citep{Bus2002a}, the 52-Color Survey \citep{Bell2005}, \citet{Rivkin2003} and \citet{deLeon2012}.
Taking into account the errors associated with superimposing spectra obtained at different epochs of observation with different equipment and the uncertainties in the spectra and in the value of the albedo ratio, the latter is not expected to exactly match the value of the spectra at 3.4 $\mu$m. 
Owing to these deviations, seen in Fig. \ref{fig:assembled}, one cannot confirm nor rule out the presence of the absorption feature based on the value of $R_p$ alone.  

Next, we combined the values of $R_p$ with the observed relative reflectances at 2.5 $\mu$m available from other datasets.  
\citet{Rivkin2003} use the parameter $1 - R_\lambda/R_{2.5}$ as a rough measure of band depth. 
Because the band minima are usually near 3.0 $\mu$m and given that W1 results from an average over 2.8 to 3.8 $\mu$m, the parameter $b \equiv 1 - R_p/R_{2.5}$ is not to be taken as a measure of band depth but as a helpful parameter to quantitatively compare the values of relative reflectances at 2.5 and 3.4 $\mu$m.
If a deep absorption band is present, $R_p$ is in general expected to be $\la R_{2.5}$ and hence $b \ga 0$, as is the case of (19) Fortuna. 
There is one general case for which this interpretation would be wrong: if the NIR slope is negative up to 3.4 $\mu$m, we would have $b \ga 0$ even if no band was present, though blue NIR slopes such as Pallas' have only been measured in a small percentage of cases \citep[e.g. 2 out of 45 in the sample of][]{deLeon2012}. 

\begin{figure}
  \centering
  \includegraphics[width=68mm]{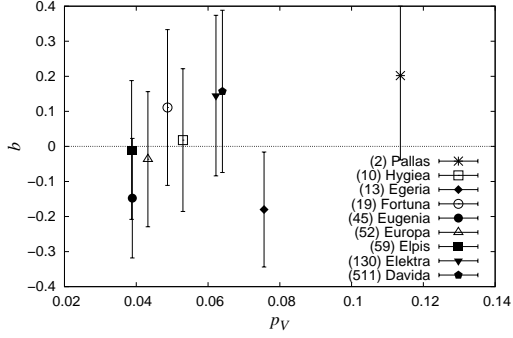}
  \caption{
    Plot of $b = 1-R_p/R_{2.5}$ versus $p_V$ for a set of primitive control objects. 
    The values of $b$ obtained from the combination of our $R_p$ and the $R_{2.5}$ taken from the assembly of spectra published by other authors (cf. figure \ref{fig:assembled}) and the shape of the 3 $\mu$m absorption band are consistent except for (13) Egeria.
  }
  \label{fig:b}
\end{figure}

A plot of $b$ vs. $p_V$ is shown in Fig. \ref{fig:b} with an estimated $b$ errorbar of 0.2 to account for the large uncertainties in $R_p$ and the assembly of the spectra.
We see that those asteroids with a weak or non-existent absorption band tend to have $b \la 0$, whereas $b \ga 0$ for those with a higher-contrast feature.
Only in the case of (13) Egeria the value of $b$ is clearly inconsistent with the absortion band observed. 
Indeed, the large errorbar in the value of $b$ still prevents irrefutable detection of the band on a case-by-case basis, but from Fig. \ref{fig:b}, the correlation between the sign of $b$ and the spectra appears robust, so that systematically obtaining $b \ga 0$ for a given population may be statistically significant. 

\begin{figure}
  \centering 
  \includegraphics[width=68mm]{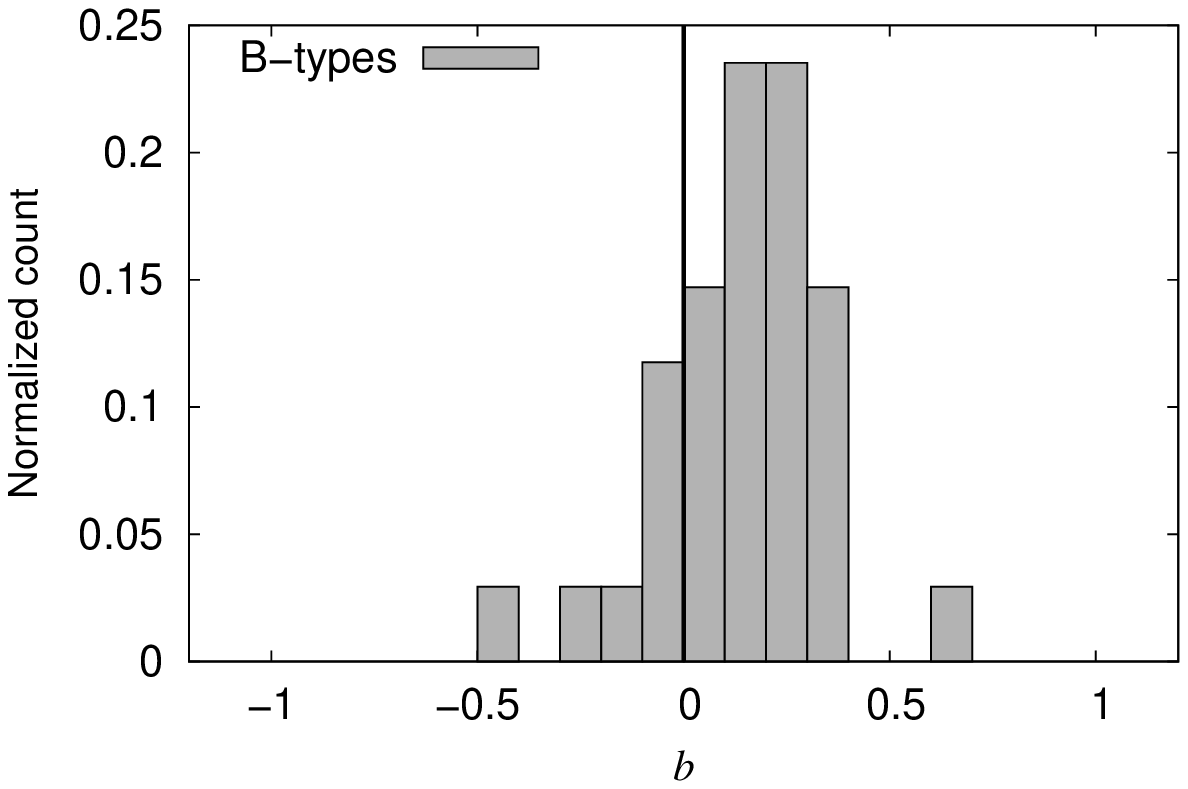}
  \includegraphics[width=68mm]{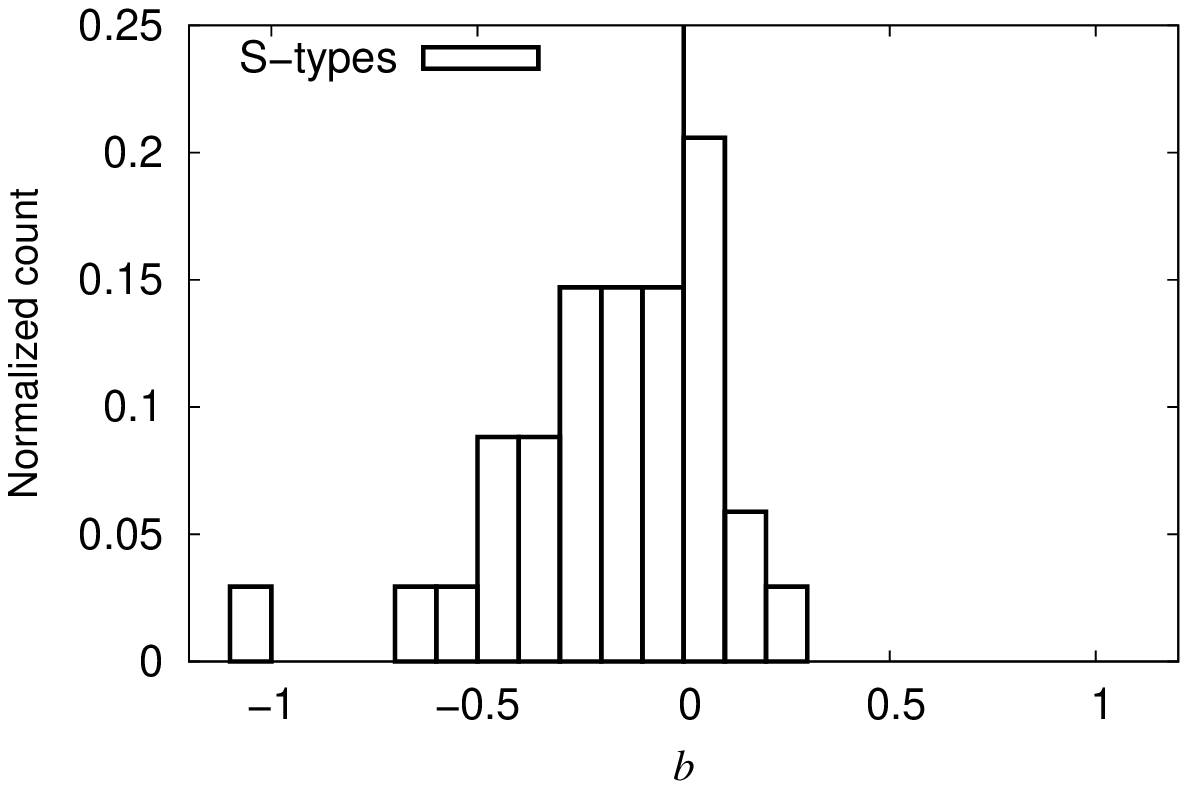}
  \caption{Normalised histograms showing the distribution of $b$-values. 
    Left panel: B-types; the fact that $b>0$ in the majority of cases suggests that a large percentage of these asteroids have absorption features in the $\sim$ 3-$\mu$m region. 
    Right panel: S-types; the opposite conclusion is reached, consistent with the anhydrous nature of S-types.
  }
  \label{fig:bsdeleon}
\end{figure}

\begin{figure}
  \centering
  \includegraphics[width=68mm]{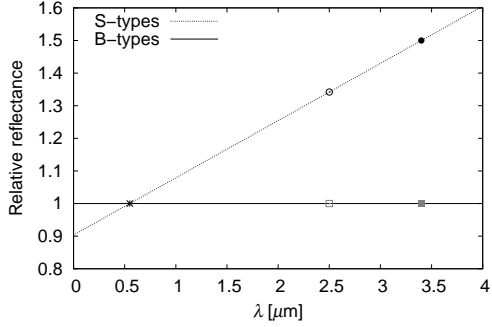}
  % \vspace{-1cm}
  \caption{
    Schematic diagram illustrating the interpolation of $R_{2.5}$ (empty points) from the average values of $R_p$ (filled points) for the B and S asteroids studied.  
  }
  \label{fig:bexplained}
\end{figure}

In Fig. \ref{fig:bsdeleon} (left panel) we show a $b$-value histogram of those B type asteroids for which the value of $b$ could be determined. 
$R_{2.5}$ was taken from the B-type spectra presented in \citet{deLeon2012}, after normalising to unity at 0.55 $\mu$m. 
In order to test for statistical significance, we resort to the Kolmogorov-Smirnov (KS) test, which enables one to reject the null hypothesis that a given set of unbinned values is compatible with having been drawn from a given distribution function \citep[see e.g. ][]{Press1986}. 
As the null hypothesis we take a gaussian distribution with zero mean. 
This choice is based on the mean value of $b$ that we would expect considering that interpolating between $R_{0.55} = 1$ and $\bar{R}_p = 1$ gives $\bar{R}_{2.5} = 1.0 \Rightarrow \bar{b} = 0$ (see Fig. \ref{fig:bexplained}). 
The KS test  amply rules out the null hypothesis that the B-types $b$-values are drawn from a gaussian distribution of zero mean regardless of its width ($\sigma$). 

To demonstrate the robustnest of this result even further, we carried out the same procedure for a list of asteroids belonging to the S-complex, including all Bus-DeMeo pure S-types for which we found the value of $R_{2.5}$ was available from the literature and for which $R_p$ could be computed, plus enough randomly selected S subtypes to get the same number of $b$-values we derived for the B-types (see Table \ref{table:bvalues}). 
S-type asteroids are ``anhydrous'' and have positive spectral slopes up to 2.5 $\mu$m so that in the absence of a 3-$\mu$m absorption feature one would expect a negative value of $b$ if the spectral slope maintains its trend up to 3.4 $\mu$m. 
Taking into account that $\bar{R}_p \simeq 1.67$ for this sample of S-types, from the interpolated value $\bar{R}_{2.5} \simeq 1.46$ (see Table \ref{supertableStypes}\addtocounter{table}{1}), the expected mean value of $b$ would be $\bar{b} = -0.14$ (see Fig. \ref{fig:bexplained}). 
The distribution of $b$-values obtained for the S-types, shown in the right panel of Fig. \ref{fig:bsdeleon}, presents a negative mean value. 
In this case, the KS test does not rule out the null hypothesis that the $b$-values are drawn from a Gaussian distribution centered at $-0.14$ with $\sigma = 0.2$ with $p$-value $> 0.9$. 
This rules out the possibility that a systematic error is causing the $b$-values of the B-type asteroids to be $> 0$. 
\begin{table}[h]
  \begin{center}                         
    \caption{\small 
      Median and standard deviation of $b$-values for B-types and S-types
      \label{table:bvalues} 
    }
    \begin{tabular}{c c c c}
      \hline\hline                
      Group & Median & $\sigma$ & $N$  \\     
      \hline                        
      B-types &  0.2 & 0.2 & 34 \\   
      S-types & -0.2  & 0.2 & 34 \\
      \hline                                
    \end{tabular}
  \end{center} 
\end{table}

In the preceeding subsection we showed that our values of $R_p$ are systematically $\sim$10\% lower than those of \citet{Masiero2011}.
We carried out the same procedure enlarging our $R_p$ values 10\% and the conclusions still hold. 

To conclude, we find that the majority of B-types with computed $b$-values verify $b>0$ and that very few present a clearly negative value of $b$, which means that even for those B-type spectra with a positive slope in the 2.5 $\mu$m region \citep[approximately half of the objects in][]{deLeon2012} there is a reduction in the reflectivity around 3.4 $\mu$m. 
As discussed above, for asteroids of the C-complex, the 3-$\mu$m absorption feature has been attributed to hydrated minerals or water-ice. 
The presence of goethite has been also proposed as an alternative explanation for this band \citep{Beck2011}. 
Nonetheless, while other closely related minerals have been found in both meteorite and asteroid spectra, extrarrestrial goethite has never been identified within the meteorite inventory, so the possibility that putative goethite-containing asteroids never found a dynamical collisional pathway to Earth is less likely than the simpler interpretation: goethite is not present in asteroidal surfaces \citep{Jewitt2012}. 
Therefore, from the distribution of $b$-values, we conclude that most asteroids in this sample (which constitutes $\la 40\%$ of the B-type population with computed $R_p$) present this absorption and that ``water'' (be it bound or free) may be common among the B-type asteroids.

\subsection{The Pallas collisional family}
\label{sec:pallasresults}

\begin{figure*}[h!]
  \centering
  \includegraphics[width=68mm]{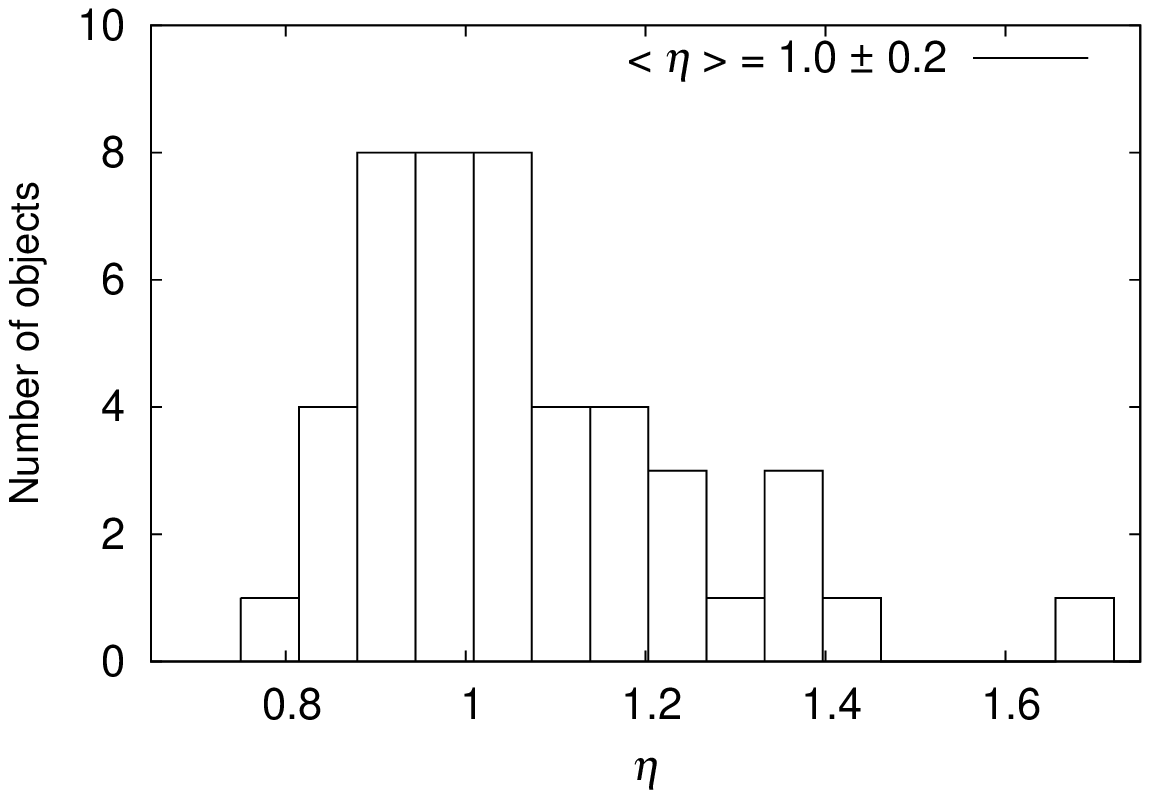}
  \includegraphics[width=68mm]{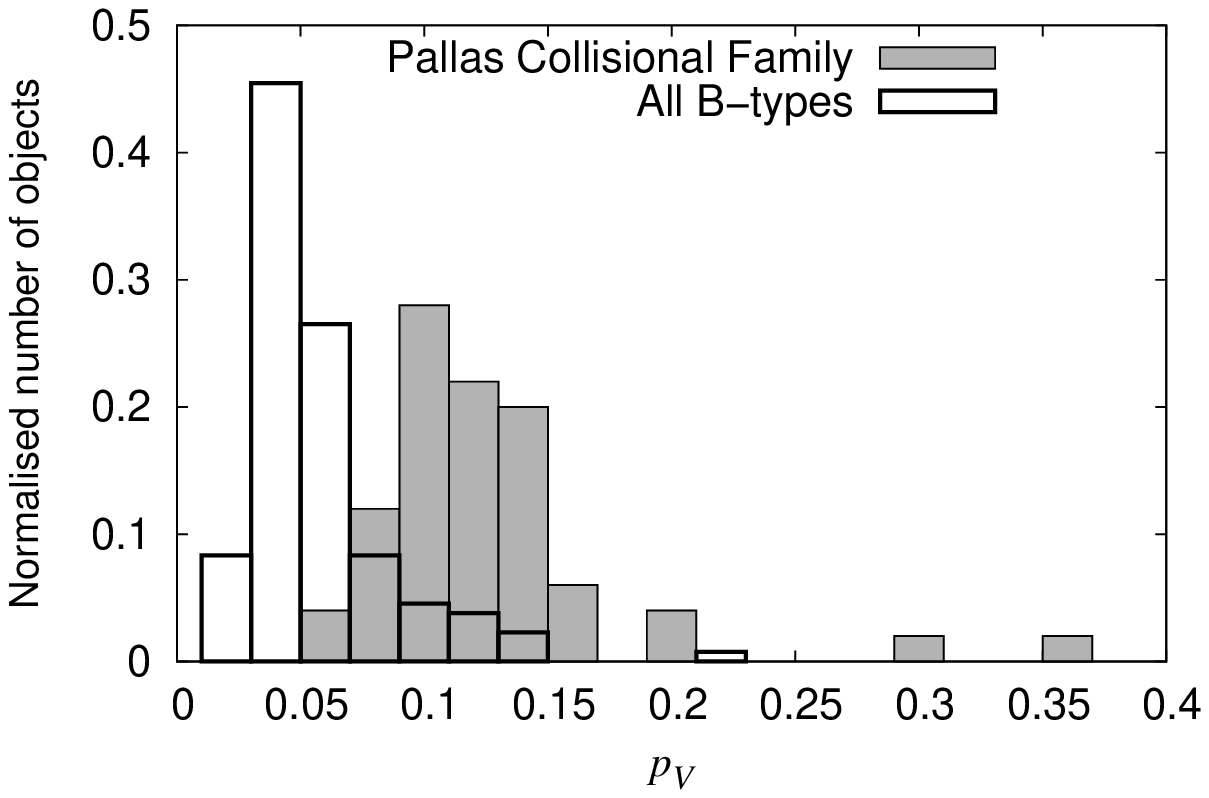}
  \caption{
    Beaming parameter (left) and geometric albedo (right) distributions of the values derived for the PCF members observed by WISE. 
    The albedo distributions are normalised, i.e. divided by their respective total number of counts. 
  } 
  \label{fig:histPallas}
\end{figure*}

We use the most up-to-date Pallas family list by \citep{Nesvorny2012}. 
WISE has observed 46 of the objects in this list. 
Histograms of beaming parameter and albedo determinations are shown in Fig. \ref{fig:histPallas} (the complete set of parameters is shown in Table \ref{supertablePallas}\addtocounter{table}{1}). 
Given that the $R_p$ value could only be fitted for seven objects, we do not include histograms for $R_p$ and $p_{IR}$. 
In Table \ref{table:meanPallas} we present the mean values of $\eta$, $p_V$ and $R_p$ along with their corresponding standard deviations and contributing number of determinations. 
\begin{table}[h!]
  \centering                         
  \caption{\small 
    Mean values and standard deviations of $\eta$, $R_p$ and $p_V$ derived for the members of the Pallas collisional family observed by WISE. 
    \label{table:meanPallas}   
  }
  \begin{tabular}{c c c c}      
    \hline\hline                
    Parameter & Mean Value & $\sigma$ & $N$  \\    
    % & & & fitted values \\ 
    \hline                        
    $\eta$ & 1.0  & 0.2  & 46 \\   
    $R_p$  & 0.5  & 0.1  & 7  \\
    $p_V$  & 0.14 & 0.05 & 50 \\
    \hline                                
  \end{tabular}
\end{table}

The average $\eta$ value of the PCF is consistent with that of the B-type population and that of the Main Belt \citep{Masiero2011}. 
However, the PCF has a moderately high albedo of $\sim$ 14\%, significantly higher than the rest of the B-types and indeed higher than the value expected for primitive asteroids, whereas their average $R_p$ is lower than the average value for the B-types (cf. Table \ref{table:mean}). 
The seven $R_p$-values computed are quite homogeneous compared to the distribution observed for the rest of the B-type population. 
As we saw in Sect. \ref{sec:parametercomp}, our parameter determinations are consistent within the errorbars with those of \citep{Masiero2011}, so we do not perform the same comparison for the PCF members. 
Masiero et al. show a histogram of $\log p_V$ with peak between $\log p_V = (-1)$ -- $(-0.8)$, i.e. $p_V = $0.10--0.16,  with which our results are consistent (they provide no mean value to compare with). 
On the other hand, though they provide the best-fit parameters in their Table 1, Masiero et al. do not discuss the $R_p$ values of the PCF.

\begin{figure*}
  \centering
  \includegraphics[width=66mm]{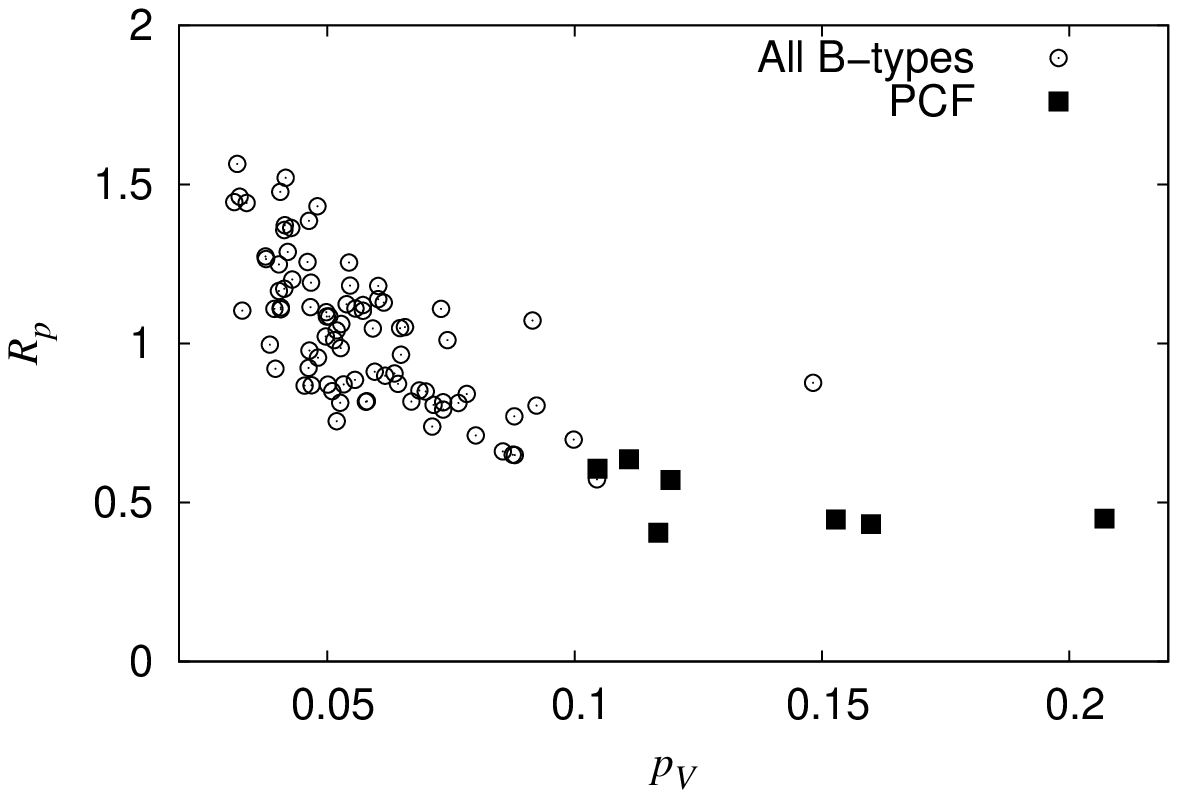}
  \includegraphics[width=66mm]{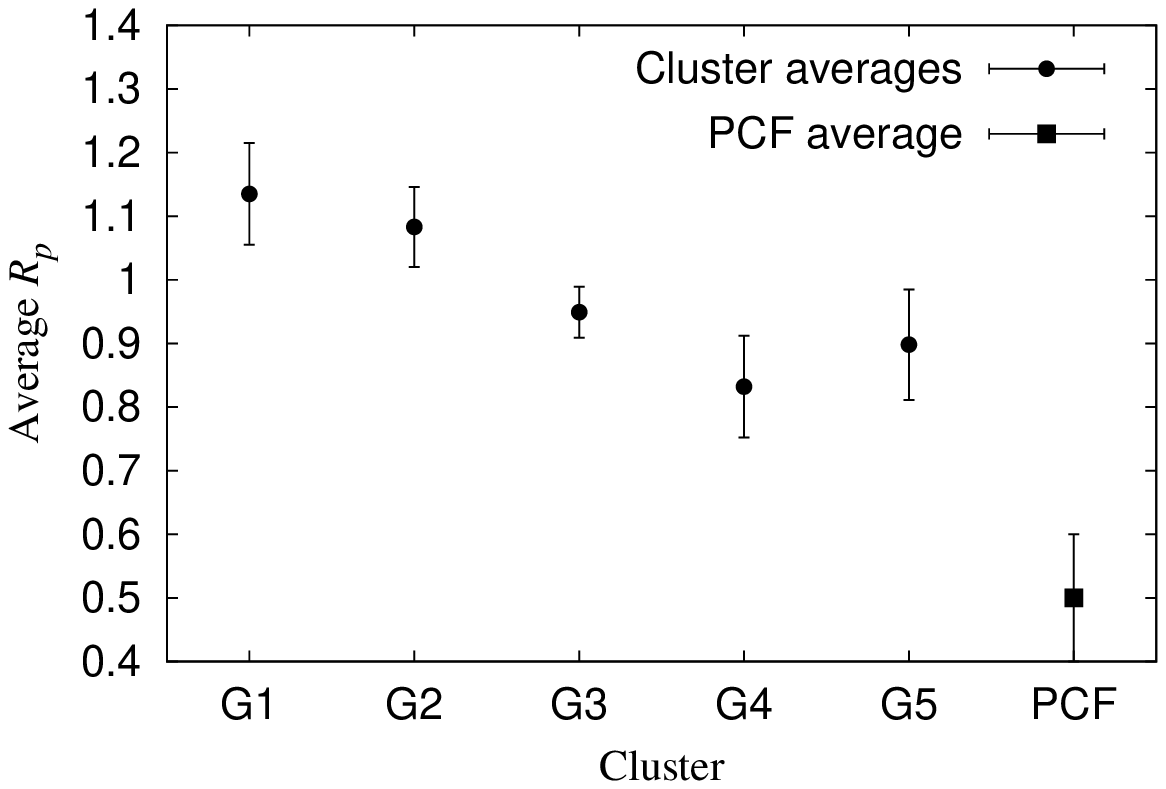}
  \caption{
    Left panel: PCF members are on average brighter in the visible and bluer in the 3.4 $\mu$m region than the B-types sample (cf. Fig. \ref{fig:rpvspvAll}). The 20\% errorbars are not plotted for clarity. Right panel: the average $R_p$ of the PCF continues the decreasing trend of the \citet{deLeon2012} clusters (cf. Fig. \ref{fig:averRpclusters}).   
  } 
  \label{fig:rpvspvPallas}
\end{figure*}

The average $p_V$ and $R_p$ values continue the trend observed for the clusters of \citet{deLeon2012} in Fig. \ref{fig:Rpclusters} and Fig. \ref{fig:averRpclusters}: they have higher visible albedos on average than the G5 cluster and are bluer at 3.4 $\mu$m than the bluest cluster, G4 (note that the average $R_p$ value of cluster G5 is obtained from only three objects, which could explain its deviation from the trend). 
Members of the Pallas-like group of objects of \citet{Clark2010} were concentrated in clusters G4 and G5 in \citet{deLeon2012} (see Sect. \ref{sec:intro}). 
In this work, the list of Pallas family members observed by WISE includes yet more asteroids. 
The four additional objects with $R_p$ determinations that were not present in \citet{deLeon2012} are also located in the lower right region of Fig. \ref{fig:Rpclusters}.
This property would be consistent with the characteristics of an extrapolated G6 centroid (see the right panel of Fig. \ref{fig:rpvspvPallas}).
Interestingly, the G6 centroid of \citet{deLeon2012} contained asteroid (3200) Phaethon alone. 
Therefore, the connection of the NEA (3200) Phaethon to the Pallas Collisional Family stablished by \citet{deLeon2010a} based on spectroscopical and dynamical arguments is also supported by the values of relative reflectances at 3.4 $\mu$m. 

\section{Discussion \label{sec:discussion}}
 
In Sect. \ref{sec:resultsRp} we present statistically significant indications that water may be common within our sample, and in Sect. \ref{sec:pallasresults} we point out how the PCF have higher geometric visible albedos and more homogeneously distributed $R_p$-values than the rest of B-types. 
The purpose of this section is to put these results in the context of other studies in the literature.

In the context of the geophysical models of the Themis and Pallas parent bodies by \citet{Castillo-Rogez2010} and \citet{Schmidt2012}, the detections of water ice and organics on the surface of (24) Themis \citep{Rivkin2010,Campins2010} and (65) Cybele \citep{Licandro2011} have been invoked as provocative indications that water has played an important role in the accretion and evolution of these asteroids' parent bodies in the mid-outer asteroid belt. 
Finding indications that a 3-$\mu$m is frequently present in B-types spectra adds further support to this conceptual framework, in which the gradual differences in the observed NIR spectral slopes of B-types might also be explained as resulting from different subsequent evolution and processing of their parent bodies. 

\citet{Schmidt2012} discuss that (1) Ceres, being the largest body of the asteroid belt and plausibly a water-rich asteroid, mostly preserved its integrity, whereas the Themis parent body was catastrophically disrupted. 
Pallas would be an intermediate case, still intact but showing evidence for heating, internal evolution, impact and loss of water. 
In accordance with this scenario, the Themis family members included in the \citet{deLeon2012} sample were distributed more or less homogeneously among the representative centroids of the complete sample, spanning from neutral to red slopes and matching several meteorite analogues, whereas the PCF members were concentrated in the bluest centroids. 
As our results show, these trends are also verified longward of 2.5 $\mu$m: the Themis family members show more heterogeneous $R_p$ values whereas the Pallas family members have distinctly higher albedos and lower, more homogeneous values of $R_p$ than the rest of B-types. 
This homogeneity is consistent with the collisional family being the result of a non-disruptive cratering event on Pallas, which has a significantly higher albedo than the rest of large B-types. 

The geophysical models by \citet{Schmidt2012} favour a water-rich past for Pallas. 
Its near spherical shape may be explained as a consequence of early melting of a substantial initial icy component that is subsequently removed by a combination of thermal and impacts processes. 
The water-loss processes that took place on the surface of Pallas may have taken place on other active bodies such as main belt comets or indeed (3200) Phaethon, linked to the Pallas collisional family by \citet{deLeon2010a}, and this activity has already been proposed to explain the characteristic NIR blue spectral slope \citep[][and references therein]{Schmidt2012}. 
According to this model, water-rich materials are also expected to be important components within some members of Pallas' family. 
The fact that the PCF members in this study are found to match the properties of the least hydrated clusters of \citet{deLeon2012} would be inconsistent with this view, but caution must be exercised when associating NIR slopes of meteorite analogues to hydration. 
For instance, while having suffered intense thermal metamorphism, CK4 chondrites (the best matches for Pallas) still show an absorption feature at 2.9 $\mu$m, albeit less prominent than e.g. CM chondrites. 

From the dynamical standpoint, the simulations of \citet{Walsh2011} also provide a congruous context for our results.  
Their model suggests that C-complex asteroids were formed in the giant-planet-forming region and that, for every C-type planetesimal from beyond 8 AU that would later be located in the outer Main Belt, $> $ 10 would have ended up in the region where terrestrial planets formed. 
Assuming that the composition of these objects is 10\% water by mass, this may account for the minimal mass required to bring the current amount of water to our planet by a factor of 6--22. 
 
On the other hand, the question of the moderately high values of $p_V$ obtained for the PCF remains unanswered. 
Finding members of the Pallas and Themis collisional families distributed differently among the \citet{deLeon2012} clusters and in Fig. \ref{fig:rpvspvPallas} (as discussed above) also leads us to ponder the possibility that a sequential or progressive physical process could explain the differences in the value distributions of $p_V$ and $R_p$.   
Unfortunately, there are many possible mechanisms underlying the observed NIR spectral variability of primitive asteroids (e.g. regolith particle sizes, space weathering as a function of asteroid familiy age and composition, thermal processing) and these are difficult to disentangle \citep{Ziffer2011}. 

\section{Conclusions \label{sec:conclusions}}

We have derived values of $D$, $\eta$, and $R_p$ of 111 B-type asteroids by means of thermal modelling of WISE data and updated $H$ values \citep[most of which have increased since the work of ][ as shown in Fig. \ref{fig:HmagDiff}]{Masiero2011}. 
Additionally, from $H$ and $D$ we have computed the corresponding values of $p_V$ (Table \ref{table:mean}). 
Our results are in agreement within the errorbars of the model with those previously published by \citet{Masiero2011}. 
However, we obtain a systematic trend of approximately $-$10\% discrepancies in the determinations of $R_p$ that we cannot explain but is most probably attributable to small differences in the tabulated solar flux data necessary to estimate the reflected light component at 3.4 $\mu$m. 
This work led to the following conclusions:
\begin{enumerate}
\item We derived the distribution of $\eta$, $p_V$ and $R_P$ fot the B-type asteroids (see Fig.  \ref{fig:histograms}) and obtained the following mean values: $\bar\eta = 1.0 \pm 0.1$, $\bar{p}_V = 0.07 \pm 0.03$, and $\bar{R}_p = 1.0 \pm 0.3$.
\item There are no high-$p_V$, high-$R_p$ B-type asteroids ($p_V > 0.10$, $R_p > 1.0$; see Fig. \ref{fig:rpvspvAll}). 
The average $R_p$-values of the centroids of \citet{deLeon2012} clearly decrease from G1 to G5, which implies a relationship between the IR slope of the asteroid spectra up to 2.5 $\mu$m (Figs. \ref{fig:Rpclusters} and \ref{fig:averRpclusters}). 
\item We computed $b$-values for a set of B-type asteroids and S-complex asteroids, which present $\bar b > 0$ and $\bar b < 0$, respectively (Fig. \ref{fig:bsdeleon}). 
While the latter result is consistent with objects of the S-complex being anhydrous, the former indicates that the majority of B-type asteroids $b$-values in this study are consistent with the presence of a 3-$\mu$m absorption feature usually attributed to hydrated minerals or water-ice; therefore, water must have played a key role in the evolution of a large fraction of the B-types, supporting recent works discussed in Sect. \ref{sec:discussion}.  
\item We have also studied the Pallas collisional family. 
On the one hand, the average albedo ($\bar{p}_V = $ 0.14 $\pm$ 0.05) of this familly is significantly higher than the average albedo of B-types ($\bar{p}_V = $ 0.07 $\pm$ 0.03) and moderately high compared to what is traditionally considered to be the albedo of primitive asteroids ($< 0.1$). 
On the other hand, the albedo ratio values of the PCF members are very low and homogeneous ($\bar{R}_p = $ 0.5 $\pm$ 0.1). 
These results clearly show the fundamental differences between the Pallas Collisional Family and the rest of B-types. 
 
In addition, the connection of the NEA (3200) Phaethon to the Pallas Collisional Family stablished by \citet{deLeon2010a} based on spectroscopical and dynamical arguments is also supported by the values of relative reflectances at 3.4 $\mu$m.  
\item Our results support the scenarios by the geophysical models by \citet{Castillo-Rogez2010} and \citet{Schmidt2012} and the simulations of \citet{Walsh2011}, which suggest that water played an important role in the origin of primitive asteroid parent bodies of the mid-outer belt. 

\end{enumerate}

\begin{acknowledgements}

  We thank the referee for a careful and constructive revision. 
  VAL acknowledges support from the project AYA2011-29489-C03-02 (MEC). 
  JL acknowledges support from the projects AYA2011-29489-C03-02 and AYA2012-39115-C03-03 (MINECO). 
  JdL thanks financial support via a "Juan de la Cierva" contract from the Spanish "Secretaría de Estado de Investigación, Desarrollo e Innovación".     
  MDB thanks the Space Situation Awareness program of the European Space Agency (ESA-SSA) for financial support.
  NPA was financed by the Spanish Ministry of Economy and Competitiveness throughout the Juan de la Cierva program. 
  
  This publication makes use of data products from NEOWISE, which is a project of the Jet Propulsion Laboratory/California Institute of Technology, funded by the Planetary Science Division of the National Aeronautics and Space Administration. 
  This research has also made use of the NASA/ IPAC Infrared Science Archive, which is operated by the Jet Propulsion Laboratory, California Institute of Technology, under contract with the National Aeronautics and Space Administration.

\end{acknowledgements}

%%%%%%%%%%%%%%%%%%%%%%%%%%%%%%%%%%%%%%%%%
% the bibliography
%%%%%%%%%%%%%%%%%%%%%%%%%%%%%%%%%%%%%%%%%

\bibliographystyle{aa}
\bibliography{/Users/victorali/jabrefdatabases/AsteroidsGeneral}

\longtab{1}{
  \begin{longtable}{c c c  c  c  c  c  c c c c c}
    \caption{\label{supertable} Best-fitting values of physical parameters determined for the B-types with WISE observations. 
      Negative values of $\eta$ and or $R_p$ indicate that the parameter was not free but fixed to the corresponding positive value. 
      Errorbars shown are minimum estimates and correspond to 10\% relative error for $D$ and 20\% for $\eta$, $p_V$ and $R_p$.}\\
    \hline\hline
    Designation & $H$ & $G$ & $D$ [km] & $p_V$ & $\eta$ & $R_p$ & W1 & W2 & W3 & W4 & $R_{2.5}$\\
    \hline
    \endfirsthead
    \caption{continued.}\\
    \hline\hline
    Designation & $H$ & $G$ & $D$ [km] & $p_V$ & $\eta$ & $R_p$ & W1 & W2 & W3 & W4 & $R_{2.5}$\\
    \hline
    \endhead
    \hline
    \endfoot
    00002 & 4.13 & 0.11 &     669 $\pm$     67 &    0.09 $\pm$   0.02 &    -1.0 $\pm$   0.2  &     0.8 $\pm$    0.2 &     3 &   3 &   0 &   0 &   0.8\\
    00024 & 7.08 & 0.19 &     187 $\pm$     19 &    0.07 $\pm$   0.01 &    -1.0 $\pm$   0.2  &     1.0 $\pm$    0.2 &     5 &   5 &   0 &   0 &   1.1\\
    00045 & 7.46 & 0.07 &     240 $\pm$     24 &    0.03 $\pm$   0.01 &    -1.0 $\pm$   0.2  &     1.6 $\pm$    0.3 &     9 &  10 &   0 &   0 &   1.3\\
    00047 & 7.84 & 0.16 &     118 $\pm$     12 &    0.09 $\pm$   0.02 &    -1.0 $\pm$   0.2  &     0.8 $\pm$    0.2 &     9 &   9 &   0 &   0 &   -\\
    00052 & 6.31 & 0.18 &     396 $\pm$     40 &    0.03 $\pm$   0.01 &    -1.0 $\pm$   0.2  &     1.4 $\pm$    0.3 &     4 &   4 &   0 &   0 &   1.3\\
    00059 & 7.93 & 0.15 &     169 $\pm$     17 &    0.04 $\pm$   0.01 &    -1.0 $\pm$   0.2  &     1.4 $\pm$    0.3 &    11 &  11 &   0 &   0 &   1.3\\
    00085 & 7.61 & 0.15 &     148 $\pm$     15 &    0.07 $\pm$   0.01 &    -1.0 $\pm$   0.2  &     1.1 $\pm$    0.2 &     6 &   6 &   0 &   0 &   1.1\\
    00141 & 8.40 & 0.15 &     129 $\pm$     13 &    0.05 $\pm$   0.01 &    -1.0 $\pm$   0.2  &     1.0 $\pm$    0.2 &     9 &   9 &   0 &   0 &   -\\
    00142 & 10.27 & 0.15 &      58 $\pm$      6 &    0.04 $\pm$   0.01 &     1.0 $\pm$  0.2  &     1.1 $\pm$    0.2 &     8 &   8 &   0 &   8 &   1.1\\
    00225 & 8.72 & 0.15 &     107 $\pm$     11 &    0.05 $\pm$   0.01 &     1.1 $\pm$   0.2  &     1.1 $\pm$    0.2 &    14 &  15 &  15 &  14 &  -\\
    00229 & 9.13 & 0.15 &     110 $\pm$     11 &    0.03 $\pm$   0.01 &     0.8 $\pm$   0.2  &     1.5 $\pm$    0.3 &    10 &  10 &   0 &  10 &  -\\
    00241 & 7.58 & 0.15 &     189 $\pm$     19 &    0.05 $\pm$   0.01 &    -1.0 $\pm$   0.2  &     1.3 $\pm$    0.3 &     7 &   7 &   0 &   0 &   -\\
    00241 & 7.58 & 0.15 &     198 $\pm$     20 &    0.04 $\pm$   0.01 &    -1.0 $\pm$   0.2  &     1.3 $\pm$    0.3 &     7 &   7 &   0 &   0 &   -\\
    00268 & 8.28 & 0.15 &     142 $\pm$     14 &    0.04 $\pm$   0.01 &    -1.0 $\pm$   0.2  &     1.4 $\pm$    0.3 &     9 &   9 &   0 &   0 &   -\\
    00282 & 10.91 & 0.15 &     41 $\pm$      4 &    0.05 $\pm$   0.01 &     1.0 $\pm$  0.2  &     -1.5 $\pm$    0.3 &     10 &  10 &   0 &  10 &  -\\
    00314 & 9.80 & 0.15 &      64 $\pm$      6 &    0.05 $\pm$   0.01 &     1.0 $\pm$   0.2  &     1.0 $\pm$    0.2 &    14 &  14 &   0 &  14 &  -\\
    00335 & 8.96 & 0.15 &      88 $\pm$      9 &    0.06 $\pm$   0.01 &     1.1 $\pm$   0.2  &     0.9 $\pm$    0.2 &     4 &   4 &   0 &   4 &   1.1\\
    00357 & 8.72 & 0.15 &     105 $\pm$     10 &    0.05 $\pm$   0.01 &     1.0 $\pm$   0.2  &     0.8 $\pm$    0.2 &     8 &   8 &   0 &   8 &   1.1\\
    00372 & 7.50 & 0.15 &     180 $\pm$     18 &    0.05 $\pm$   0.01 &    -1.0 $\pm$   0.2  &     1.3 $\pm$    0.3 &    12 &  12 &   0 &   0 &   -\\
    00379 & 8.87 & 0.15 &      88 $\pm$      9 &    0.06 $\pm$   0.01 &     1.0 $\pm$   0.2  &     1.0 $\pm$    0.2 &     6 &   6 &   0 &   6 &   1.2\\
    00383 & 9.91 & 0.15 &      44 $\pm$      4 &    0.10 $\pm$   0.02 &     1.3 $\pm$   0.2  &     0.7 $\pm$    0.1 &    12 &  13 &  13 &  13 &  0.9\\
    00400 & 10.50 & 0.15 &      39 $\pm$      4 &    0.07 $\pm$   0.01 &     1.2 $\pm$  0.2  &     0.8 $\pm$    0.2 &    11 &  11 &  11 &  10 &  -\\
    00404 & 9.01 & 0.15 &      98 $\pm$     10 &    0.05 $\pm$   0.01 &     1.0 $\pm$   0.2  &     0.9 $\pm$    0.2 &     8 &   8 &   0 &   8 &  -\\
    00426 & 8.42 & 0.15 &     117 $\pm$     12 &    0.06 $\pm$   0.01 &     0.9 $\pm$   0.2  &     1.1 $\pm$    0.2 &     9 &   9 &   0 &   6 &   0.8\\
    00431 & 8.72 & 0.15 &     103 $\pm$     10 &    0.05 $\pm$   0.01 &     0.9 $\pm$   0.2  &     1.1 $\pm$    0.2 &    13 &  12 &   0 &  12 &  1.2\\
    00461 & 10.48 & 0.15 &      46 $\pm$      5 &    0.05 $\pm$   0.01 &     1.0 $\pm$  0.2  &     1.1 $\pm$    0.2 &    26 &  27 &  27 &  27 &  -\\
    00464 & 9.52 & 0.15 &      84 $\pm$      8 &    0.04 $\pm$   0.01 &     1.0 $\pm$   0.2  &     1.1 $\pm$    0.2 &     4 &   4 &   0 &   4 &  -\\
    00464 & 9.52 & 0.15 &      82 $\pm$      8 &    0.04 $\pm$   0.01 &     0.9 $\pm$   0.2  &     1.2 $\pm$    0.2 &    26 &  26 &   0 &  26 &  -\\
    00468 & 9.83 & 0.15 &      66 $\pm$      7 &    0.05 $\pm$   0.01 &     1.0 $\pm$   0.2  &     1.2 $\pm$    0.2 &    13 &  13 &   0 &  13 &  -\\
    00526 & 10.17 & 0.15 &      48 $\pm$      5 &    0.06 $\pm$   0.01 &     1.1 $\pm$  0.2  &     0.9 $\pm$    0.2 &    13 &  12 &  17 &  17 &  -\\
    00531 & 12.00 & 0.15 &      16 $\pm$      2 &    0.10 $\pm$   0.02 &     0.9 $\pm$  0.2  &    -1.5 $\pm$    0.3 &     0 &   8 &  12 &  12 &  -\\
    00541 & 10.10 & 0.15 &      57 $\pm$      6 &    0.05 $\pm$   0.01 &     1.0 $\pm$  0.2  &     1.0 $\pm$    0.2 &    13 &  11 &   0 &  13 &  -\\
    00555 & 10.70 & 0.15 &      33 $\pm$      3 &    0.09 $\pm$   0.02 &     1.0 $\pm$  0.2  &     0.7 $\pm$    0.1 &    12 &  12 &  12 &  12 &  -\\
    00560 & 10.90 & 0.15 &      36 $\pm$      4 &    0.06 $\pm$   0.01 &     1.0 $\pm$  0.2  &     0.8 $\pm$    0.2 &    14 &  14 &   9 &  14 &  -\\
    00567 & 9.16 & 0.15 &      91 $\pm$      9 &    0.05 $\pm$   0.01 &     1.0 $\pm$   0.2  &     1.4 $\pm$    0.3 &     8 &   8 &   0 &   7 &   -\\
    00567 & 9.16 & 0.15 &      82 $\pm$      8 &    0.06 $\pm$   0.01 &     1.0 $\pm$   0.2  &     1.1 $\pm$    0.2 &     8 &   8 &   0 &   8 &   -\\
    00635 & 9.01 & 0.15 &      97 $\pm$     10 &    0.05 $\pm$   0.01 &     0.9 $\pm$   0.2  &     0.9 $\pm$    0.2 &    11 &  11 &   0 &   8 &   -\\
    00702 & 7.25 & 0.15 &     202 $\pm$     20 &    0.05 $\pm$   0.01 &    -1.0 $\pm$   0.2  &     1.2 $\pm$    0.2 &    10 &  10 &   0 &   0 &   -\\
    00704 & 5.94 & -0.02 &     361 $\pm$     36 &    0.06 $\pm$   0.01 &    -1.0 $\pm$   0.2  &     1.1 $\pm$   0.2 &    6  &   7 &   0 &   0 &   -\\
    00704 & 5.94 & -0.02 &     351 $\pm$     35 &    0.06 $\pm$   0.01 &    -1.0 $\pm$   0.2  &     1.1 $\pm$   0.2 &    8  &   8 &   0 &   0 &   -\\
    00762 & 8.28 & 0.15 &     144 $\pm$     14 &    0.04 $\pm$   0.01 &    -1.0 $\pm$   0.2  &     1.4 $\pm$    0.3 &    10 &  10 &   0 &   0 &   -\\
    00767 & 10.10 & 0.15 &      47 $\pm$      5 &    0.07 $\pm$   0.01 &     1.1 $\pm$  0.2  &     0.8 $\pm$    0.2 &     9 &  12 &  12 &  12 &  1.0\\
    00893 & 9.47 & 0.15 &      76 $\pm$      8 &    0.05 $\pm$   0.01 &     0.9 $\pm$   0.2  &     1.1 $\pm$    0.2 &     8 &   9 &   0 &   9 &   -\\
    00895 & 8.20 & 0.15 &     123 $\pm$     12 &    0.06 $\pm$   0.01 &     1.1 $\pm$   0.2  &     0.9 $\pm$    0.2 &     7 &   7 &   0 &   7 &   -\\
    00954 & 9.94 & 0.15 &      52 $\pm$      5 &    0.07 $\pm$   0.01 &     1.1 $\pm$   0.2  &     0.9 $\pm$    0.2 &    12 &  12 &   0 &  12 &  -\\
    00981 & 10.57 & 0.15 &      34 $\pm$      3 &    0.09 $\pm$   0.02 &     1.1 $\pm$  0.2  &     0.6 $\pm$    0.1 &     7 &   8 &   8 &   8 &   -\\
    00988 & 11.60 & 0.15 &      22 $\pm$      2 &    0.08 $\pm$   0.02 &     0.9 $\pm$  0.2  &    -1.5 $\pm$    0.3 &     0 &   0 &  10 &  10 &  -\\
    00988 & 11.60 & 0.15 &      22 $\pm$      2 &    0.09 $\pm$   0.02 &     0.9 $\pm$  0.2  &    -1.5 $\pm$    0.3 &     0 &   0 &  10 &  10 &  -\\
    01003 & 10.70 & 0.15 &      36 $\pm$      4 &    0.07 $\pm$   0.01 &     1.3 $\pm$  0.2  &     0.8 $\pm$    0.2 &     8 &   8 &   8 &   8 &  1.0\\
    01003 & 10.70 & 0.15 &      34 $\pm$      3 &    0.08 $\pm$   0.02 &     1.4 $\pm$  0.2  &     0.8 $\pm$    0.2 &    12 &   8 &  13 &  13 &  1.0\\
    01021 & 8.98 & 0.15 &     105 $\pm$     11 &    0.04 $\pm$   0.01 &     1.0 $\pm$   0.2  &     1.1 $\pm$    0.2 &    12 &  12 &   0 &  12 &  1.0\\
    01035 & 10.20 & 0.15 &      60 $\pm$      6 &    0.04 $\pm$   0.01 &     1.0 $\pm$  0.2  &     1.3 $\pm$    0.2 &    12 &  14 &  14 &  14 &  1.4\\
    01076 & 12.30 & 0.15 &      24 $\pm$      2 &    0.04 $\pm$   0.01 &     1.0 $\pm$  0.2  &     1.0 $\pm$    0.2 &     5 &   7 &   8 &   8 &  1.0\\
    01076 & 12.30 & 0.15 &      23 $\pm$      2 &    0.04 $\pm$   0.01 &     0.9 $\pm$  0.2  &     0.9 $\pm$    0.2 &    13 &  13 &  13 &  13 &  1.0\\
    01109 & 10.06 & 0.15 &      64 $\pm$      6 &    0.04 $\pm$   0.01 &     0.9 $\pm$  0.2  &     1.5 $\pm$    0.3 &    11 &  11 &   0 &  11 &  -\\
    01109 & 10.06 & 0.15 &      63 $\pm$      6 &    0.04 $\pm$   0.01 &     0.9 $\pm$  0.2  &     1.5 $\pm$    0.3 &     7 &   7 &   0 &   7 &   -\\
    01154 & 10.51 & 0.15 &      59 $\pm$      6 &    0.03 $\pm$   0.01 &     1.0 $\pm$  0.2  &     1.4 $\pm$    0.3 &    11 &  12 &   9 &  12 &  -\\
    01213 & 11.10 & 0.15 &      31 $\pm$      3 &    0.06 $\pm$   0.01 &     1.1 $\pm$  0.2  &     1.0 $\pm$    0.2 &    12 &  13 &  15 &  15 &  -\\
    01229 & 11.30 & 0.15 &      30 $\pm$      3 &    0.06 $\pm$   0.01 &     1.1 $\pm$  0.2  &     1.2 $\pm$    0.2 &     7 &   0 &  13 &  13 &  -\\
    01331 & 10.14 & 0.15 &      39 $\pm$      4 &    0.10 $\pm$   0.02 &     1.1 $\pm$  0.2  &     0.6 $\pm$    0.1 &    13 &  14 &  14 &  14 &  -\\
    01340 & 11.10 & 0.15 &      32 $\pm$      3 &    0.06 $\pm$   0.01 &     1.2 $\pm$  0.2  &     0.9 $\pm$    0.2 &     7 &   9 &   9 &   9 &  -\\
    01362 & 11.18 & 0.15 &      30 $\pm$      3 &    0.07 $\pm$   0.01 &     1.1 $\pm$  0.2  &    -1.5 $\pm$    0.3 &     0 &   0 &  12 &  12 &  -\\
    01444 & 11.30 & 0.15 &      28 $\pm$      3 &    0.07 $\pm$   0.01 &     1.0 $\pm$  0.2  &    -1.5 $\pm$    0.3 &     0 &   0 &  15 &  15 &  -\\
    01474 & 12.66 & 0.15 &      15 $\pm$      2 &    0.06 $\pm$   0.01 &     0.8 $\pm$  0.2  &    -1.5 $\pm$    0.3 &     0 &   0 &   8 &   8 &   -\\
    01474 & 12.66 & 0.15 &      15 $\pm$      2 &    0.07 $\pm$   0.01 &     0.8 $\pm$  0.2  &    -1.5 $\pm$    0.3 &     0 &   0 &   6 &   6 &   -\\
    01484 & 10.80 & 0.15 &      41 $\pm$      4 &    0.05 $\pm$   0.01 &     0.9 $\pm$  0.2  &     1.0 $\pm$    0.2 &    11 &  11 &   0 &  11 &  1.5\\
    01493 & 11.99 & 0.15 &      23 $\pm$      2 &    0.05 $\pm$   0.01 &     0.9 $\pm$  0.2  &     -1.5 $\pm$    0.3 &    17 &  17 &   0 &  17 &  -\\
    01508 & 12.03 & 0.15 &      16 $\pm$      2 &    0.11 $\pm$   0.02 &     1.0 $\pm$  0.2  &    -1.5 $\pm$    0.3 &     0 &   0 &   5 &   5 &   -\\
    01539 & 11.10 & 0.15 &      26 $\pm$      3 &    0.09 $\pm$   0.02 &     1.0 $\pm$  0.2  &     1.1 $\pm$    0.2 &     8 &   0 &  12 &  12 &  1.1\\
    01576 & 11.04 & 0.15 &      30 $\pm$      3 &    0.08 $\pm$   0.02 &     1.2 $\pm$  0.2  &     0.8 $\pm$    0.2 &     8 &   8 &   8 &   8 &   -\\
    01579 & 10.68 & 0.15 &      50 $\pm$      5 &    0.04 $\pm$   0.01 &     0.9 $\pm$  0.2  &     1.2 $\pm$    0.3 &    9  &  10 &  12 &  12 &  -\\
    01615 & 11.38 & 0.15 &      31 $\pm$      3 &    0.05 $\pm$   0.01 &     1.1 $\pm$  0.2  &     0.9 $\pm$    0.2 &    11 &  13 &  12 &  12 &  -\\
    01655 & 11.04 & 0.15 &      40 $\pm$      4 &    0.04 $\pm$   0.01 &     1.0 $\pm$  0.2  &     1.2 $\pm$    0.2 &    11 &  12 &  12 &  12 &  1.4\\
    01655 & 11.04 & 0.15 &      38 $\pm$      4 &    0.05 $\pm$   0.01 &     1.0 $\pm$  0.2  &     1.1 $\pm$    0.2 &    11 &  11 &  11 &  11 &  1.4\\
    01693 & 10.97 & 0.15 &      39 $\pm$      4 &    0.05 $\pm$   0.01 &     0.9 $\pm$  0.2  &     1.0 $\pm$    0.2 &    11 &  11 &   0 &  11 &  -\\
    01705 & 13.20 & 0.15 &      11 $\pm$      1 &    0.07 $\pm$   0.01 &     1.0 $\pm$  0.2  &    -1.5 $\pm$    0.3 &     0 &   6 &   9 &   9 &   -\\
    01705 & 13.20 & 0.15 &      13 $\pm$      1 &    0.06 $\pm$   0.01 &     1.1 $\pm$  0.2  &    -1.5 $\pm$    0.3 &     0 &  27 &  26 &  27 &  -\\
    01724 & 11.30 & 0.15 &      40 $\pm$      4 &    0.03 $\pm$   0.01 &     1.0 $\pm$  0.2  &     1.1 $\pm$    0.2 &     6 &  10 &  10 &  10 &  -\\
    01768 & 12.70 & 0.15 &      21 $\pm$      2 &    0.03 $\pm$   0.01 &     1.0 $\pm$  0.2  &    -1.5 $\pm$    0.3 &     0 &   8 &   8 &   8 &   -\\
    01768 & 12.70 & 0.15 &      20 $\pm$      2 &    0.04 $\pm$   0.01 &     1.0 $\pm$  0.2  &    -1.5 $\pm$    0.3 &     0 &  12 &  12 &  12 &  -\\
    01796 & 9.84 & 0.15 &       71 $\pm$      7 &    0.04 $\pm$   0.01 &     0.9 $\pm$   0.2  &    1.2 $\pm$    0.2 &    17 &  16 &   0 &  16 &  -\\
    01901 & 11.40 & 0.15 &      27 $\pm$      3 &    0.07 $\pm$   0.01 &     1.0 $\pm$  0.2  &     1.1 $\pm$    0.2 &     8 &  10 &  11 &  11 &  -\\
    02096 & 13.50 & 0.15 &      12 $\pm$      1 &    0.05 $\pm$   0.01 &     1.3 $\pm$  0.2  &    -1.5 $\pm$    0.3 &     0 &   0 &   5 &   5 &   -\\
    02332 & 10.60 & 0.15 &      36 $\pm$      4 &    0.08 $\pm$   0.02 &     1.1 $\pm$  0.2  &     0.9 $\pm$    0.1 &    10 &  11 &  11 &  11 &  1.3\\
    02332 & 10.60 & 0.15 &      34 $\pm$      3 &    0.09 $\pm$   0.02 &     1.1 $\pm$  0.2  &     0.9 $\pm$    0.1 &    10 &  13 &  13 &  13 &  1.3\\
    02446 & 12.90 & 0.15 &      13 $\pm$      1 &    0.07 $\pm$   0.01 &     1.2 $\pm$  0.2  &     0.8 $\pm$    0.2 &     6 &  10 &  10 &  10 &  1.2\\
    02446 & 12.90 & 0.15 &      15 $\pm$      2 &    0.05 $\pm$   0.01 &     1.2 $\pm$  0.2  &     0.9 $\pm$    0.2 &    12 &  13 &  13 &  13 &  1.2\\
    02464 & 11.70 & 0.15 &      23 $\pm$      2 &    0.07 $\pm$   0.01 &     1.1 $\pm$  0.2  &    -1.5 $\pm$    0.3 &     0 &   7 &  14 &  14 &  -\\
    02519 & 11.50 & 0.15 &      22 $\pm$      2 &    0.09 $\pm$   0.02 &     1.1 $\pm$  0.2  &    -1.5 $\pm$    0.3 &     0 &   0 &  13 &  13 &  -\\
    02524 & 11.10 & 0.15 &      35 $\pm$      4 &    0.05 $\pm$   0.01 &     1.0 $\pm$  0.2  &     1.0 $\pm$    0.2 &    10 &  10 &  10 &  10 &  -\\
    02525 & 10.90 & 0.15 &      33 $\pm$      3 &    0.07 $\pm$   0.01 &     1.1 $\pm$  0.2  &     0.8 $\pm$    0.2 &    11 &  12 &  12 &  11 &  -\\
    02629 & 14.90 & 0.15 &       5 $\pm$      1 &    0.07 $\pm$   0.01 &     1.2 $\pm$  0.2  &     0.7 $\pm$    0.1 &    15 &  19 &  19 &  19 &  -\\
    02659 & 11.60 & 0.15 &      29 $\pm$      3 &    0.05 $\pm$   0.01 &     1.0 $\pm$  0.2  &     1.3 $\pm$    0.3 &     8 &   9 &  11 &  11 &  -\\
    02708 & 12.00 & 0.15 &      22 $\pm$      2 &    0.06 $\pm$   0.01 &     1.2 $\pm$  0.2  &     1.0 $\pm$    0.2 &    7  &  13 &  14 &  14 &  1.4\\
    02772 & 13.60 & 0.15 &      10 $\pm$      1 &    0.07 $\pm$   0.01 &     0.9 $\pm$  0.2  &    -1.5 $\pm$    0.3 &     0 &  12 &  13 &  12 &  -\\
    02809 & 13.60 & 0.15 &      12 $\pm$      1 &    0.04 $\pm$   0.01 &     1.0 $\pm$  0.2  &    -1.5 $\pm$    0.3 &     0 &  12 &  13 &  13 &  -\\
    02973 & 12.90 & 0.15 &      15 $\pm$      1 &    0.06 $\pm$   0.01 &     0.9 $\pm$  0.2  &    -1.5 $\pm$    0.3 &     0 &  11 &  11 &  11 &  -\\
    02973 & 12.90 & 0.15 &      12 $\pm$      1 &    0.08 $\pm$   0.02 &     0.9 $\pm$  0.2  &    -1.5 $\pm$    0.3 &     0 &   9 &  11 &  11 &  -\\
    03000 & 13.60 & 0.15 &      11 $\pm$      1 &    0.05 $\pm$   0.01 &     1.1 $\pm$  0.2  &    -1.5 $\pm$    0.3 &     0 &   0 &   6 &   6 &   -\\
    03000 & 13.60 & 0.15 &      11 $\pm$      1 &    0.06 $\pm$   0.01 &     1.0 $\pm$  0.2  &    -1.5 $\pm$    0.3 &     0 &   0 &  15 &  15 &  -\\
    03036 & 10.30 & 0.15 &      47 $\pm$      5 &    0.06 $\pm$   0.01 &     1.0 $\pm$  0.2  &     1.0 $\pm$    0.2 &     9 &  10 &  10 &  10 &  1.3\\
    03036 & 10.30 & 0.15 &      51 $\pm$      5 &    0.05 $\pm$   0.01 &     1.0 $\pm$  0.2  &     1.0 $\pm$    0.2 &     8 &   8 &   8 &   8 &   1.3\\
    03074 & 13.60 & 0.15 &      10 $\pm$      1 &    0.06 $\pm$   0.01 &     1.2 $\pm$  0.2  &    -1.5 $\pm$    0.3 &     0 &   5 &   9 &   9 &   -\\
    03139 & 10.70 & 0.15 &      41 $\pm$      4 &    0.06 $\pm$   0.01 &     1.1 $\pm$  0.2  &     0.9 $\pm$    0.2 &    10 &  12 &  11 &  10 &  -\\
    03162 & 11.30 & 0.15 &      31 $\pm$      3 &    0.05 $\pm$   0.01 &     1.1 $\pm$  0.2  &    -1.5 $\pm$    0.2 &    13 &   8 &  16 & 16  &  1.3\\
    03162 & 11.30 & 0.15 &      33 $\pm$      3 &    0.05 $\pm$   0.01 &     1.1 $\pm$  0.2  &     0.9 $\pm$    0.2 &    10 &  10 &  10 &  11 &  1.3\\
    03204 & 12.20 & 0.15 &      21 $\pm$      2 &    0.05 $\pm$   0.01 &     1.2 $\pm$  0.2  &    -1.5 $\pm$    0.3 &     0 &   0 &   7 &   7 &   -\\
    03204 & 12.20 & 0.15 &      20 $\pm$      2 &    0.06 $\pm$   0.01 &     1.3 $\pm$  0.2  &    -1.5 $\pm$    0.3 &     0 &   0 &   7 &   7 &   -\\
    03566 & 12.90 & 0.15 &      14 $\pm$      1 &    0.07 $\pm$   0.01 &     0.9 $\pm$  0.2  &    -1.5 $\pm$    0.3 &     0 &   5 &   8 &   8 &   -\\
    03579 & 13.60 & 0.15 &       7 $\pm$      1 &    0.12 $\pm$   0.02 &     0.9 $\pm$  0.2  &     0.4 $\pm$    0.1 &    10 &  14 &  13 &  13 &  1.0\\
    03579 & 13.60 & 0.15 &       7 $\pm$      1 &    0.13 $\pm$   0.03 &     1.0 $\pm$  0.2  &    -1.5 $\pm$    0.3 &     0 &   0 &   9 &   9 &  1.0\\
    03581 & 12.10 & 0.15 &      14 $\pm$      1 &    0.13 $\pm$   0.03 &     0.9 $\pm$  0.2  &    -1.5 $\pm$    0.3 &     0 &   0 &  13 &  13 &  -\\
    03627 & 13.50 & 0.15 &      11 $\pm$      1 &    0.06 $\pm$   0.01 &     1.0 $\pm$  0.2  &    -1.5 $\pm$    0.3 &     0 &  10 &  10 &  10 &  -\\
    03647 & 11.50 & 0.15 &      28 $\pm$      3 &    0.06 $\pm$   0.01 &     1.0 $\pm$  0.2  &     0.8 $\pm$    0.2 &    11 &  12 &  12 &  12 &  -\\
    04100 & 11.50 & 0.15 &      17 $\pm$      2 &    0.15 $\pm$   0.03 &     1.0 $\pm$  0.2  &     0.9 $\pm$    0.2 &     5 &   0 &  11 &  11 &  1.1\\
    04396 & 13.60 & 0.15 &       5 $\pm$      1 &    0.24 $\pm$   0.05 &     1.1 $\pm$  0.2  &    -1.5 $\pm$    0.3 &     0 &   0 &   8 &   8 &  -\\
    04484 & 12.30 & 0.15 &      16 $\pm$      2 &    0.08 $\pm$   0.02 &     1.1 $\pm$  0.2  &    -1.5 $\pm$    0.3 &     0 &   6 &   9 &   9 &  -\\
    04837 & 11.60 & 0.15 &      28 $\pm$      3 &    0.05 $\pm$   0.01 &     0.9 $\pm$  0.2  &    -1.5 $\pm$    0.3 &     0 &   0 &  10 &  10 &  -\\
    04955 & 11.70 & 0.15 &      22 $\pm$      2 &    0.08 $\pm$   0.02 &     1.2 $\pm$  0.2  &    -1.5 $\pm$    0.3 &     0 &   0 &  12 &  12 &  -\\
    04997 & 12.70 & 0.15 &      10 $\pm$      1 &    0.16 $\pm$   0.03 &     1.1 $\pm$  0.2  &    -1.5 $\pm$    0.3 &     0 &   0 &  10 &  10 &  -\\
    05057 & 12.30 & 0.15 &      18 $\pm$      2 &    0.06 $\pm$   0.01 &     1.0 $\pm$  0.2  &    -1.5 $\pm$    0.3 &     0 &   0 &   8 &   8 &  -\\
    05133 & 11.90 & 0.15 &      24 $\pm$      2 &    0.05 $\pm$   0.01 &     1.1 $\pm$  0.2  &    -1.5 $\pm$    0.3 &     0 &   8 &  12 &  12 &  -\\
    05222 & 11.30 & 0.15 &      19 $\pm$      2 &    0.15 $\pm$   0.03 &     1.0 $\pm$  0.2  &     0.4 $\pm$    0.1 &    12 &  13 &  13 &  13 &  -\\
    05234 & 11.90 & 0.15 &      15 $\pm$      1 &    0.15 $\pm$   0.03 &     1.0 $\pm$  0.2  &    -1.5 $\pm$    0.3 &     0 &   0 &   6 &   6 &  -\\
    05234 & 11.90 & 0.15 &     14 $\pm$      2 &    0.13 $\pm$   0.03 &     0.9 $\pm$  0.2  &     -1.5 $\pm$    0.3 &     0 &   0 &  10 &  10 &  -\\
    05330 & 12.10 & 0.15 &     15 $\pm$      2 &    0.11 $\pm$   0.02 &     0.9 $\pm$  0.2  &      0.6 $\pm$    0.1 &     9 &  16 &  16 &  15 &  -\\
    05870 & 13.20 & 0.15 &      9 $\pm$      1 &    0.11 $\pm$   0.02 &     1.0 $\pm$  0.2  &     -1.5 $\pm$    0.3 &     0 &   0 &   5 &   4 &  -\\
    06297 & 12.30 & 0.15 &     18 $\pm$      2 &    0.07 $\pm$   0.01 &     1.1 $\pm$  0.2  &     -1.5 $\pm$    0.3 &     0 &   0 &   8 &   8 &  -\\
    08518 & 12.80 & 0.15 &     14 $\pm$      1 &    0.07 $\pm$   0.01 &     1.1 $\pm$  0.2  &     -1.5 $\pm$    0.3 &     0 &   0 &   5 &   5 &  -\\
    08519 & 13.60 & 0.15 &      7 $\pm$      1 &    0.12 $\pm$   0.02 &     1.2 $\pm$  0.2  &     -1.5 $\pm$    0.3 &     0 &   0 &  11 &  11 &  -\\
    08906 & 12.80 & 0.15 &     14 $\pm$      1 &    0.07 $\pm$   0.01 &     0.9 $\pm$  0.2  &     -1.5 $\pm$    0.3 &     0 &   0 &   6 &   6 &  -\\
    09219 & 11.90 & 0.15 &     20 $\pm$      2 &    0.07 $\pm$   0.01 &     1.3 $\pm$  0.2  &     -1.5 $\pm$    0.3 &     0 &   0 &   8 &   8 &  -\\
  \end{longtable}                                                                                                                              
}                                                                                                                                            

\longtab{3}{
  \begin{longtable}{c c c  c  c  c  c  c c c c c}
    \caption{\label{supertableStypes} Best-fitting values of physical parameters determined for the S-types with sufficient W1 WISE observations and published 2.5 $\mu$m reflectances. 
      Negative values of $\eta$ and or $R_p$ indicate that the parameter was not free but fixed to the corresponding positive value. 
      Errorbars shown are minimum estimates and correspond to 10\% relative error for $D$ and 20\% for $\eta$, $p_V$ and $R_p$.}\\
    \hline\hline
    Designation & $H$ & $G$ & $D$ [km] & $p_V$ & $\eta$ & $R_p$ & W1 & W2 & W3 & W4 & $R_{2.5}$\\
    \hline
    \endfirsthead
    \caption{continued.}\\
    \hline\hline
    Designation & $H$ & $G$ & $D$ [km] & $p_V$ & $\eta$ & $R_p$ & W1 & W2 & W3 & W4 & $R_{2.5}$\\
    \hline
    \endhead
    \hline
    \endfoot                                                                                                                                           
    00009  &  6.280  &  0.170  &  231  $\pm$  23  &  0.10  $\pm$  0.02  &  -1.0  $\pm$  0.2  &  2.1  $\pm$  0.4  &  8  &  8 &  0  &  0  &  1.5     \\
    00009  &  6.280  &  0.170  &  188  $\pm$  19  &  0.15  $\pm$  0.03  &  -1.0  $\pm$  0.2  &  1.9  $\pm$  0.4  &  13  &  13 &  0  &  0  &  1.5   \\
    00011  &  6.550  &  0.150  &  171  $\pm$  17  &  0.14 $\pm$  0.03  &  -1.0  $\pm$  0.2  &  1.6  $\pm$  0.3  &  8  &  8 &  0  &  0  &  1.4       \\
    00017  &  7.760  &  0.150  &  93  $\pm$  9  &  0.16  $\pm$  0.03  &  0.9  $\pm$  0.2  &  1.8  $\pm$  0.4  &  10  &  10 &  0  &  10  &  1.5     \\
    00026  &  7.400  &  0.150  &  86  $\pm$  9  &  0.26  $\pm$  0.05  &  1.1  $\pm$  0.2  &  1.3  $\pm$  0.3  &  11  &  12 &  0  &  12  &  1.4     \\
    00029  &  5.850  &  0.200  &  194  $\pm$  19  &  0.21  $\pm$  0.04  &  1.0  $\pm$  0.2  &  1.4  $\pm$  0.3  &  10  &  10 &  0  &  0  &  1.3    \\
    00030  &  7.570  &  0.150  &  88  $\pm$  9  &  0.21  $\pm$  0.04  &  -1.0  $\pm$  0.2  &  1.8  $\pm$  0.4  &  11  &  11 &  0  &  0  &  1.5     \\
    00032  &  7.560  &  0.150  &  78  $\pm$  8  &  0.27  $\pm$  0.05  &  -1.0  $\pm$  0.2  &  1.4  $\pm$  0.3  &  7  &  7 &  0  &  0  &  1.4        \\
    00043  &  7.930  &  0.110  &  76  $\pm$  8  &  0.21  $\pm$  0.04  &  1.2  $\pm$  0.2  &  1.7  $\pm$  0.3  &  3  &  3 &  0  &  3  &  1.3         \\
    00057  &  7.030  &  0.150  &  114  $\pm$  11  &  0.21  $\pm$  0.04  &  -1.0  $\pm$  0.2  &  1.6  $\pm$  0.3  &  9  &  9 &  0  &  0  &  1.4     \\
    00061  &  7.680  &  0.150  &  93  $\pm$  9  &  0.17  $\pm$  0.04  &  -1.0  $\pm$  0.2  &  1.8  $\pm$  0.3  &  10  &  10 &  0  &  0  &  1.3     \\
    00079  &  7.960  &  0.250  &  71  $\pm$  7  &  0.23  $\pm$  0.05  &  1.1  $\pm$  0.2  &  1.3  $\pm$  0.3  &  11  &  11 &  0  &  11  &  1.3      \\
    00103  &  7.660  &  0.150  &  87  $\pm$  9  &  0.20  $\pm$  0.045  &  -1.0  $\pm$  0.2  &  1.6  $\pm$  0.3  &  9  &  9 &  0  &  0  &  1.5      \\
    00119  &  8.420  &  0.150  &  65  $\pm$  7  &  0.18  $\pm$  0.04  &  1.2  $\pm$  0.2  &  1.9  $\pm$  0.4  &  10  &  10 &  0  &  10  &  1.6     \\
    00151  &  9.100  &  0.150  &  43  $\pm$  4  &  0.22  $\pm$  0.04  &  1.1  $\pm$  0.2  &  1.4  $\pm$  0.3  &  20  &  21 &  0  &  21  &  1.7      \\
    00158  &  9.270  &  0.150  &  44  $\pm$  4  &  0.18  $\pm$  0.04  &  1.3  $\pm$  0.3  &  1.5  $\pm$  0.3  &  15  &  15 &  15  &  15  &  1.7    \\
    00192  &  7.130  &  0.030  &  99  $\pm$  10  &  0.25  $\pm$  0.05  &  1.2  $\pm$  0.2  &  1.5 $\pm$  0.3  &  8  &  8 &  0  &  8  &  1.6         \\
    00192  &  7.130  &  0.030  &  99  $\pm$  10  &  0.25  $\pm$  0.05  &  1.2  $\pm$  0.2  &  1.5  $\pm$  0.3  &  6  &  6 &  0  &  5  &  1.6        \\
    00245  &  7.820  &  0.150  &  80  $\pm$  8  &  0.21  $\pm$  0.04  &  1.0  $\pm$  0.2  &  1.6  $\pm$  0.3  &  7  &  7 &  0  &  7  &  1.3         \\
    00288  &  9.840  &  0.150  &  33  $\pm$  3  &  0.19  $\pm$  0.04  &  0.9  $\pm$  0.2  &  1.4  $\pm$  0.3  &  9  &  10 &  11  &  11  &  1.3     \\
    00371  &  8.720  &  0.150  &  59  $\pm$  6  &  0.17  $\pm$  0.03  &  1.1  $\pm$  0.2  &  1.7  $\pm$  0.3  &  9  &  10 &  0  &  10  &  1.3      \\
    00532  &  5.810  &  0.260  &  193  $\pm$  19  &  0.22  $\pm$  0.05  &  -1.0  $\pm$  0.2  &  1.1  $\pm$  0.2  &  12  &  12 &  0  &  0  &  1.3    \\
    00584  &  8.710  &  0.240  &  54  $\pm$  5  &  0.20  $\pm$  0.04  &  1.0  $\pm$  0.2  &  3.2 $\pm$  0.7  &  5  &  4 &  0  &  6  &  1.6          \\
    00631  &  8.700  &  0.150  &  52  $\pm$  5  &  0.22  $\pm$  0.04  &  1.1  $\pm$  0.2  &  1.4  $\pm$  0.3  &  21  &  21 &  0  &  21  &  1.4     \\
    00699  &  11.720  &  0.150  &  13  $\pm$  1  &  0.23  $\pm$  0.05  &  1.0  $\pm$  0.2  &  2.0  $\pm$  0.4  &  4  &  0 &  8  &  8  &  1.2        \\
    00793  &  10.260  &  0.150  &  29  $\pm$  3  &  0.16  $\pm$  0.03  &  1.1  $\pm$  0.2  &  1.7  $\pm$  0.3  &  9  &  9 &  9  &  9  &  1.1       \\
    00793  &  10.260  &  0.150  &  29  $\pm$  3  &  0.16  $\pm$  0.03 &  1.1  $\pm$  0.2  &  1.7  $\pm$  0.3  &  13  &  13 &  13  &  13  &  1.1    \\
    00847  &  10.290  &  0.150  &  29  $\pm$  3  &  0.16  $\pm$  0.03  &  1.0  $\pm$  0.2  &  1.6 $\pm$  0.3  &  15  &  15 &  15  &  15  &  1.4    \\
    01036  &  9.450  &  0.300  &  36  $\pm$  4  &  0.22  $\pm$  0.04  &  1.0  $\pm$  0.2  &  1.2  $\pm$  0.3  &  9  &  9 &  9  &  9  &  1.3         \\
    01036  &  9.450  &  0.300  &  38  $\pm$  4  &  0.20  $\pm$  0.04  &  1.  $\pm$  0.2  &  1.3  $\pm$  0.3  &  13  &  13 &  13  &  13  &  1.3      \\
    01866  &  12.400  &  0.150  &  8  $\pm$  1  &  0.29  $\pm$  0.06  &  1.1  $\pm$  0.2  &  1.3  $\pm$  0.3  &  7  &  0 &  10  &  10  &  1.6       \\
    01980  &  13.920  &  0.150  &  5  $\pm$  1  &  0.16  $\pm$  0.03  &  1.4  $\pm$  0.3  &  2.7  $\pm$  0.5  &  5  &  0 &  7  &  6  &  1.7         \\
    01980  &  13.920  &  0.150  &  6  $\pm$  1  &  0.15  $\pm$  0.03  &  1.4  $\pm$  0.3  &  2.1  $\pm$  0.4  &  9  &  9 &  10  &  10  &  1.7       \\
    11500  &  18.400  &  0.150  &  0.8  $\pm$  0.1  &  0.12  $\pm$  0.02  &  1.6  $\pm$  0.3  &  1.7  $\pm$  0.3  &  10  &  20 &  20  &  20  &  1.4  \\
  \end{longtable}                                                                                                                              
}

\longtab{4}{
  \begin{longtable}{c c c  c  c  c  c  c c c c}
    \caption{\label{supertablePallas} Best-fitting values of physical parameters determined for the Pallas Collisional Family asteroids excluding (2) Pallas with WISE observations. 
      Negative values of $\eta$ and or $R_p$ indicate that the parameter was not free but fixed to the corresponding positive value. 
      Errorbars shown are minimum estimates and correspond to 10\% relative error for $D$, and 20\% for $\eta$, $p_V$ and $R_p$.}\\
    \hline\hline
    Designation & $H$ & $G$ & $D$ [km] & $p_V$ & $\eta$ & $R_p$ & W1 & W2 & W3 & W4\\
    \hline
    \endfirsthead
    \caption{continued.}\\
    \hline\hline
    Designation & $H$ & $G$ & $D$ [km] & $p_V$ & $\eta$ & $R_p$ & W1 & W2 & W3 & W4\\
    \hline
    \endhead
    \hline
    \endfoot
    00531 & 12.00 & 0.15 &    16 $\pm$    2 &    0.10 $\pm$   0.02 &     0.9 $\pm$    0.2 &    -1.5 $\pm$    0.3 &    0 &   8 &  12 &  12 \\
    03579 & 13.60 & 0.15 &     7 $\pm$    1 &    0.12 $\pm$   0.02 &     0.9 $\pm$    0.2 &     0.4 $\pm$    0.1 &   13 &  14 &  13 &  13 \\
    03579 & 13.60 & 0.15 &     7 $\pm$    1 &    0.13 $\pm$   0.03 &     1.0 $\pm$    0.2 &    -1.5 $\pm$    0.3 &    0 &   0 &   9 &   9 \\
    05222 & 11.30 & 0.15 &    19 $\pm$    2 &    0.15 $\pm$   0.03 &     1.0 $\pm$    0.2 &     0.4 $\pm$    0.1 &   13 &  13 &  13 &  13 \\
    05234 & 11.90 & 0.15 &    14 $\pm$    1 &    0.15 $\pm$   0.03 &     1.0 $\pm$    0.2 &    -1.5 $\pm$    0.3 &    0 &   0 &   6 &   6 \\
    05234 & 11.90 & 0.15 &    15 $\pm$    2 &    0.13 $\pm$   0.03 &     0.9 $\pm$    0.2 &    -1.5 $\pm$    0.3 &    0 &   0 &  10 &  10 \\
    05330 & 12.10 & 0.15 &    15 $\pm$    2 &    0.11 $\pm$   0.02 &     0.9 $\pm$    0.2 &     0.6 $\pm$    0.1 &   10 &  16 &  16 &  15 \\
    08009 & 13.70 & 0.15 &     6 $\pm$    1 &    0.14 $\pm$   0.03 &     1.3 $\pm$    0.2 &    -1.5 $\pm$    0.3 &    0 &   0 &  11 &   9 \\
    11064 & 12.60 & 0.15 &     9 $\pm$    1 &    0.21 $\pm$   0.04 &     1.2 $\pm$    0.2 &     0.4 $\pm$    0.1 &   16 &  17 &  17 &  17 \\
    12377 & 12.60 & 0.15 &    10 $\pm$    1 &    0.15 $\pm$   0.03 &     0.9 $\pm$    0.2 &    -1.5 $\pm$    0.3 &    0 &   0 &   9 &   9 \\
    14916 & 13.50 & 0.15 &     8 $\pm$    1 &    0.12 $\pm$   0.02 &     0.8 $\pm$    0.2 &    -1.5 $\pm$    0.3 &    0 &   0 &   8 &   9 \\
    15834 & 13.20 & 0.15 &     9 $\pm$    1 &    0.11 $\pm$   0.02 &     0.8 $\pm$    0.2 &    -1.5 $\pm$    0.3 &    0 &   0 &   9 &   9 \\
    23830 & 13.50 & 0.15 &     10 $\pm$    1 &    0.07 $\pm$   0.01 &     1.0 $\pm$    0.2 &    -1.5 $\pm$    0.3 &    0 &   5 &  12 &  12 \\
    24793 & 13.80 & 0.15 &     7 $\pm$    1 &    0.11 $\pm$   0.02 &     1.2 $\pm$    0.2 &    -1.5 $\pm$    0.3 &    0 &   0 &  12 &  12 \\
    24793 & 13.80 & 0.15 &     7 $\pm$    1 &    0.13 $\pm$   0.03 &     1.1 $\pm$    0.2 &    -1.5 $\pm$    0.3 &    0 &  10 &  13 &  13 \\
    25853 & 13.30 & 0.15 &     8 $\pm$    1 &    0.14 $\pm$   0.03 &     1.0 $\pm$    0.2 &    -1.5 $\pm$    0.3 &    0 &   0 &   6 &   6 \\
    33166 & 12.90 & 0.15 &    10 $\pm$    1 &    0.12 $\pm$   0.02 &     1.3 $\pm$    0.2 &    -1.5 $\pm$    0.3 &    0 &   0 &  14 &  14 \\
    33750 & 12.50 & 0.15 &    12 $\pm$    1 &    0.12 $\pm$   0.02 &     1.0 $\pm$    0.2 &     0.6 $\pm$    0.1 &   14 &  14 &  14 &  14 \\
    36273 & 12.80 & 0.15 &     9 $\pm$    1 &    0.18 $\pm$   0.04 &     1.0 $\pm$    0.2 &    -1.5 $\pm$    0.3 &    0 &  16 &  17 &  17 \\
    39646 & 13.50 & 0.15 &     4.4 $\pm$    0.4 &    0.36 $\pm$   0.07 &     0.8 $\pm$    0.2 &    -1.5 $\pm$    0.3 &    0 &   0 &  10 &  12 \\
    40101 & 14.10 & 0.15 &     6 $\pm$    1 &    0.10 $\pm$   0.02 &     1.4 $\pm$    0.2 &    -1.5 $\pm$    0.3 &    0 &   0 &   6 &   3 \\
    44232 & 13.10 & 0.15 &     9 $\pm$    1 &    0.12 $\pm$   0.02 &     1.1 $\pm$    0.2 &    -1.5 $\pm$    0.3 &    0 &  14 &  14 &  14 \\
    46037 & 13.70 & 0.15 &     6 $\pm$    1 &    0.15 $\pm$   0.03 &     1.1 $\pm$    0.2 &    -1.5 $\pm$    0.3 &    0 &   0 &   8 &   8 \\
    52229 & 13.60 & 0.15 &     8 $\pm$    1 &    0.09 $\pm$   0.02 &     1.0 $\pm$    0.2 &    -1.5 $\pm$    0.3 &    0 &  11 &  11 &  11 \\
    57050 & 13.50 & 0.15 &     7 $\pm$    1 &    0.17 $\pm$   0.03 &     0.8 $\pm$    0.2 &    -1.5 $\pm$    0.3 &    0 &   0 &   6 &   4 \\
    66714 & 14.30 & 0.15 &     6 $\pm$    1 &    0.09 $\pm$   0.02 &     1.1 $\pm$    0.2 &    -1.5 $\pm$    0.3 &    0 &   7 &   7 &   7 \\
    66714 & 14.30 & 0.15 &     5.5 $\pm$    0.5 &    0.11 $\pm$   0.02 &     1.2 $\pm$    0.2 &    -1.5 $\pm$    0.3 &    0 &   0 &   7 &   7 \\
    66803 & 12.50 & 0.15 &     8 $\pm$    1 &    0.32 $\pm$   0.06 &     1.0 $\pm$    0.2 &    -1.5 $\pm$    0.3 &    0 &   0 &  15 &  15 \\
    67370 & 13.70 & 0.15 &     6 $\pm$    1 &    0.15 $\pm$   0.03 &     1.0 $\pm$    0.2 &    -1.5 $\pm$    0.3 &    0 &   0 &  11 &  12 \\
    67779 & 12.60 & 0.15 &    10 $\pm$    1 &    0.16 $\pm$   0.03 &     0.9 $\pm$    0.2 &     0.4 $\pm$    0.1 &   11 &  13 &  13 &  13 \\
    69371 & 13.70 & 0.15 &     7 $\pm$    1 &    0.11 $\pm$   0.02 &     1.0 $\pm$    0.2 &    -1.5 $\pm$    0.3 &    0 &   0 &  21 &  20 \\
    69931 & 13.90 & 0.15 &     7 $\pm$    1 &    0.11 $\pm$   0.02 &     0.9 $\pm$    0.2 &    -1.5 $\pm$    0.3 &    0 &  12 &  12 &  12 \\
    82899 & 13.60 & 0.15 &     7 $\pm$    1 &    0.13 $\pm$   0.03 &     1.1 $\pm$    0.2 &    -1.5 $\pm$    0.3 &    0 &   0 &  11 &  11 \\
    87006 & 13.90 & 0.15 &     6 $\pm$    1 &    0.14 $\pm$   0.03 &     1.0 $\pm$    0.2 &    -1.5 $\pm$    0.3 &    0 &   7 &   9 &   9 \\
    90368 & 13.40 & 0.15 &     7 $\pm$    1 &    0.14 $\pm$   0.03 &     1.0 $\pm$    0.2 &    -1.5 $\pm$    0.3 &    0 &   0 &   8 &   8 \\
    A0590 & 14.60 & 0.15 &    5.1 $\pm$  0.5 &    0.10 $\pm$   0.02 &     1.2 $\pm$    0.2 &    -1.5 $\pm$    0.3 &    0 &   0 &  20 &  18 \\
    A1283 & 13.90 & 0.15 &     6 $\pm$    1 &    0.15 $\pm$   0.03 &     0.9 $\pm$    0.2 &    -1.5 $\pm$    0.3 &    0 &   0 &   7 &   7 \\
    A3779 & 14.40 & 0.15 &     3.9 $\pm$    0.4 &    0.21 $\pm$   0.04 &    -1.0 $\pm$    0.2 &    -1.5 $\pm$    0.3 &    0 &   0 &   8 &   0 \\
    A9640 & 13.70 & 0.15 &     8 $\pm$    1 &    0.10 $\pm$   0.02 &     1.1 $\pm$    0.2 &     0.6 $\pm$    0.1 &   17 &  21 &  21 &  21 \\
    B3770 & 14.10 & 0.15 &     6 $\pm$    1 &    0.11 $\pm$   0.02 &     1.0 $\pm$    0.2 &    -1.5 $\pm$    0.3 &    0 &   0 &  11 &  11 \\
    B8223 & 14.40 & 0.15 &     4.7 $\pm$    0.5 &    0.14 $\pm$   0.03 &     1.4 $\pm$    0.2 &    -1.5 $\pm$    0.3 &    0 &   0 &  13 &  12 \\
    C3349 & 14.70 & 0.15 &     4.1 $\pm$    0.4 &    0.14 $\pm$   0.03 &     1.2 $\pm$    0.2 &    -1.5 $\pm$    0.3 &    0 &  10 &  14 &  14 \\
    D6038 & 15.00 & 0.15 &     5.0 $\pm$    0.5 &    0.07 $\pm$   0.01 &    -1.0 $\pm$    0.2 &    -1.5 $\pm$    0.3 &    0 &   0 &   5 &   0 \\
    D8406 & 14.80 & 0.15 &     4.6 $\pm$    0.5 &    0.10 $\pm$   0.02 &     0.9 $\pm$    0.2 &    -1.5 $\pm$    0.3 &    0 &   0 &   9 &   8 \\
    E5861 & 14.60 & 0.15 &     4.5 $\pm$    0.5 &    0.13 $\pm$   0.03 &    -1.0 $\pm$    0.2 &    -1.5 $\pm$    0.3 &    0 &   0 &   3 &   0 \\
    F7914 & 14.10 & 0.15 &     7 $\pm$    1 &    0.09 $\pm$   0.02 &     1.2 $\pm$    0.2 &    -1.5 $\pm$    0.3 &    0 &   0 &   7 &   3 \\
    H6413 & 14.70 & 0.15 &     4.4 $\pm$    0.4 &    0.12 $\pm$   0.02 &     1.7 $\pm$    0.2 &    -1.5 $\pm$    0.3 &    0 &   0 &   7 &   5 \\
    I8324 & 14.50 & 0.15 &     6 $\pm$    1 &    0.09 $\pm$   0.02 &     1.3 $\pm$    0.2 &    -1.5 $\pm$    0.3 &    0 &   0 &   8 &   4 \\
    K6956 & 14.30 & 0.15 &     4.8 $\pm$    0.5 &    0.15 $\pm$   0.03 &     1.3 $\pm$    0.2 &    -1.5 $\pm$    0.3 &    0 &   0 &   4 &   3 \\
    N4076 & 14.20 & 0.15 &     4.7 $\pm$    0.5 &    0.17 $\pm$   0.03 &    -1.0 $\pm$    0.2 &    -1.5 $\pm$    0.3 &    0 &   0 &   4 &   0 \\
  \end{longtable}
}

\appendix
\section{Thermal modelling of WISE asteroid data \label{app:model}}

Our aim is to model the observed asteroid flux as a function of a number of physical parameters and derive the set of parameter values that most closely reproduce the actually measured fluxes. 
In this work we follow the method described by \citet{Mainzer2011b}.
The set of wavelengths covered by WISE (specified in Sect. \ref{sec:data}) allow us to derive up to three parameters by fitting a thermal model to asteroid WISE data: asteroid effective diameter, beaming parameter and reflectance at 3.4 $\mu$m (defined below).
Within the wavelength range covered, the observed asteroid flux consists of two components: 
\begin{equation}
  F^{(m)}_\lambda = f_{\mathrm{th},\lambda} + r_{\mathrm{s},\lambda}.
  \label{ec:modelflux}
\end{equation}
The thermal flux component ($f_{\mathrm{th},\lambda}$) is the main contribution to W3 and W4 whereas the reflected sunlight component ($r_{\mathrm{s},\lambda}$) dominates in band W1.
In general, W2 will have non-negligible contributions from both components \citep{Mainzer2011b}.

The computation of $f_{\mathrm{th},\lambda}$ is based on the Near Earth Asteroid Thermal Model \citep[NEATM; see][]{Harris1998,Delbo2002}.
The asteroid is assumed to be spherical and its surface is divided into triangular facets which contribute to the total thermal flux observed by WISE in accordance to the facet temperature ($T_i$), the geocentric distance ($\Delta$) and the phase angle ($\alpha_\odot$).  
In turn, the temperature of each facet depends on the asteroid heliocentric distance ($r_\odot$) and its orientation with respect to the direction towards the sun. It is given by
\begin{equation}
  \frac{S_\odot}{r^2_\odot}(1-A)\mu_i\delta a_i = \eta\sigma\epsilon T_i^4\delta a_i\,,
  \label{ec:model_thermo}
\end{equation}
which results from assuming that each surface element $\delta a_i$ is in instantaneous equilibrium with solar radiation.
 $S_\odot$ is the solar power at a distance of 1 AU, $A$ is the bolometric Bond albedo, $\epsilon$ is the emissivity \citep[usually taken to be 0.9; see][and references therein]{Delbo2007}, $\sigma$ is the Stefan-Boltzmann constant and $\mu_i = \cos\theta_i$, where $\theta_i$ is the angle between the normal to the surface element $i$ and the direction towards the sun.
Non-illuminated facets will be instantaneously in equilibrium with the very low temperatures of the surroundings ($\sim$ 0 K) and thus their contribution to $f_{\mathrm{th},\lambda}$ is neglected in the NEATM.
Finally, the beaming parameter ($\eta$) can be thought of as a normalisation or calibration factor that accounts for the different effects that would change the apparent day-side temperature distribution of the asteroid compared to that of a perfectly smooth, non-rotating sphere \citep{Harris1998}. 
These include, for example, the enhanced sunward thermal emission due to surface roughness ($\eta < 1$), or the non-negligible night-side emission of surfaces with high thermal inertia which, in order to conserve energy, causes the day-side temperature to be lower than that compared to the ideal case with zero thermal inertia ($\eta > 1$).

The asteroid thermal flux component is then given by
\begin{equation}
  f_{\mathrm{th},\lambda} = \Omega\sum_if_{i,\lambda}(T_i)\,,
  \label{ec:tfc}
\end{equation}
where $f_{i,\lambda}$ is the contribution from each illuminated facet of a 1-km sphere; 
$\Omega\equiv (D/1 \,\mathrm{km})^2$ scales the cross-section of the latter to the corresponding value of an asteroid of diameter $D$. 
 The color correction associated with each value of $T_i$ and each WISE band is applied to the facet flux. 
 By definition, it is the quotient of the in-band flux of the black-body at the given temperature to that of Vega \citep{Wright2010}. 
 A color correction table was generated for all integer temperatures from 70 K up to 1000 K using the filter profiles available from \citet{Cutri2012}. 
%The files contain the RSR as function of $\lambda$ in steps of 0.075 \AA  except for the longer wavelengths, which were omitted}.

The reflected light component, the second term in the right hand side of Eq. \ref{ec:modelflux}, is calculated as follows. 
First, the asteroid visible magnitude ($V$) that would be observed at a given geometry ($r_\odot$, $\Delta$ and $\alpha_\odot$) can be estimated using the IAU phase curve correction \citep{Bowell1989}, along with the tabulated values of asteroid absolute magnitude ($H$) and slope parameter ($G$) from the Minor Planet Center.
Secondly, knowledge of the solar visible magnitude and flux at 0.55 $\mu$m ($V_\odot$ and $f_{V_\odot}$, respectively) allows us to calculate the sunlight reflected from the asteroid at that particular wavelength:
\begin{equation}
  r_V = f_{V_\odot}\times 10^{-\frac{V - V_\odot}{2.5}}.
  \label{ec:rlc_fv}
\end{equation}
If we assume that the sun is well approximated by a black body emitter at the solar effective temperature ($T_\odot = 5778 $ K), the estimated reflected flux at any other desired wavelength ($r_\lambda$) can be computed by normalising the black body emission $B_\lambda(T_\odot)$ to verify $r_V$, i.e. 
\begin{equation}
r_\lambda = r_V \frac{B_\lambda(T_\odot)}{B_V(T_\odot)}.
\end{equation}
In this approximation, we can also consider
\begin{equation}
\frac{B_{IR}(T_\odot)}{B_V(T_\odot)} \approx \frac{f_{IR_\odot}}{f_{V_\odot}},
\label{ec:rlc_approx}
\end{equation}
from which we arrive at the following expression:
\begin{equation}
  r_\lambda = f_{IR_\odot} \frac{B_\lambda(T_\odot)}{B_{IR}(T_\odot)} \times 10^{-\frac{V - V_\odot}{2.5}},
  \label{ec:rlc_rlambda}
\end{equation}
where the subscript $IR$ denotes 3.4 $\mu$m.
Note that we do not color correct this component given the small correction to the flux of a G2V star \cite[see Table 1 of][]{Wright2010}. 
Finally, in order to account for possible differences in the reflectivity at wavelengths longward of 0.55 $\mu$m, a prefactor to $r_{\lambda}$ is included in the model, such that
\begin{equation}
r_{\mathrm{s},\lambda} = R_pr_{\lambda}.
\label{ec:rlc_appendix}
\end{equation}
This prefactor, $R_p$, is by definition equivalent to the ratio of $p_{IR}$ and the the visible geometric albedo, so we will refer to it as the ``albedo ratio''. 
The paremeter $p_{IR}$ is the reflectivity at 3.4 and 4.6 $\mu$m defined by \citet{Mainzer2011b}. 
%This procedure actually assumes that the same factor applies to all WISE bands, but its effect on W3 and W4 is negligible since $r_{\lambda} \ll f_{\mathrm{th},\lambda}$ at these wavelengths.

To sum up, the observed model flux can then be written as: 
\begin{equation}
F^{(m)}_\lambda = \Omega\sum_if_{i,\lambda}\left[T_i(\eta)\right] + R_pr_\lambda.
\label{ec:finalmodel}
\end{equation}
We use the Levenberg-Marquardt algorithm \citep{Press1986} in order to find the values of asteroid size ($D = \sqrt{\Omega}$, in km), beaming parameter ($\eta$) and albedo ratio ($R_p$) that minimise the $\chi^2$ of the asteroid's WISE data set, namely
\begin{equation}
\chi^2 = \sum_{j,\lambda}\left(\frac{F_{j,\lambda} - F^{(m)}_{j,\lambda}}{\sigma_{j,\lambda}}\right)^2,
\label{ec:chi2}
\end{equation}
where $F_{j,\lambda}$ and $\sigma_{j,\lambda}$ are the measured fluxes and corresponding uncertainties, $j$ runs over the observation epochs and $\lambda$ labels the WISE bands.
The implementation of this technique involves the calculation of the partial derivatives of $F^{(m)}_\lambda$ with respect to the fitting parameters, which is straightforward in the case of $\Omega$ and $R_p$.
The partial derivative with respect to $\eta$ can be derived from:
\begin{equation}
\frac{\partial F^{(m)}_\lambda}{\partial \eta} = \left(\frac{\partial F^{(m)}_\lambda}{\partial T_i}\right)\left(\frac{\partial T_i}{\partial \eta}\right).
\label{ec:chainrule}
\end{equation}
The derivation of the exact analytical expresion for the first factor is considerably long to be included here, whereas the derivative of $T_i$ with respect to $\eta$ is easily obtained from Eq. \ref{ec:model_thermo}.

%The number of parameters we can fit for each object depends upon how many and which WISE bands are present in its data set.
%Whenever there is one single or no thermal band available (W2, W3 or W4) we assume $\eta=1$.0; $R_p$ is fitted only if we have W1 data, and it is fixed to 1.5 otherwise. 
%These default values are based on the peak of their respective fitted value distributions of Main Belt Asteroids presented in \citet{Masiero2011}.
%Nonetheless, if no W1 data is used, the value of $R_p$ plays no effective role in the fits as long as W2 is dominated by thermal emission, which is usually the case for the asteroids in our sample.

\section{Comparison with \citet{Masiero2011} \label{app:paramcomp}}

\begin{figure*}[t]
  \begin{center}
    \includegraphics[width=68mm]{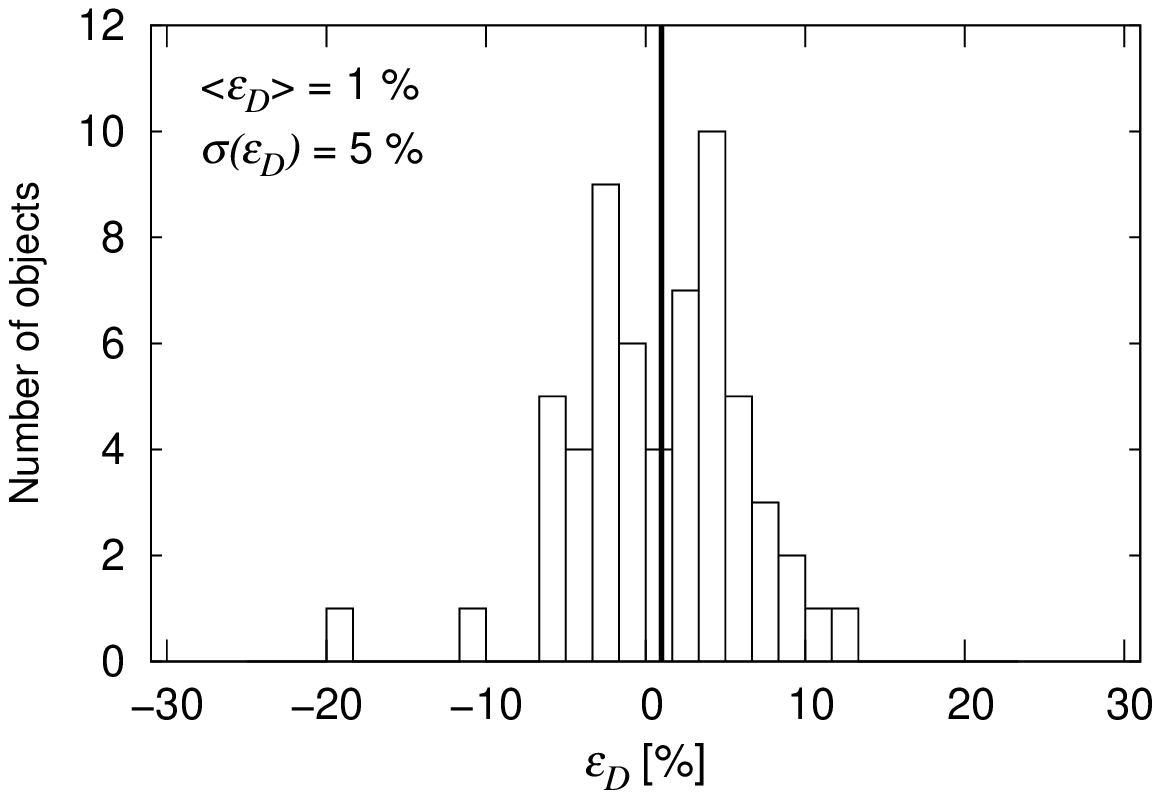}
    \includegraphics[width=68mm]{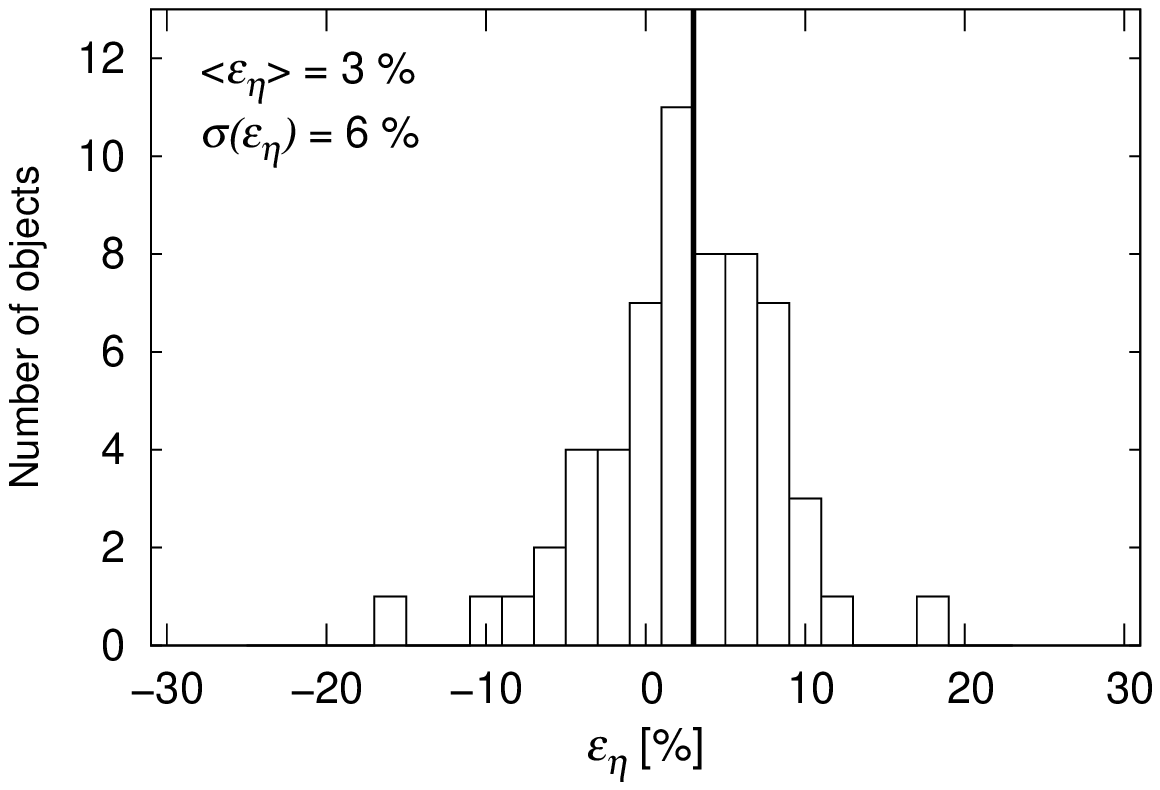}

    \includegraphics[width=68mm]{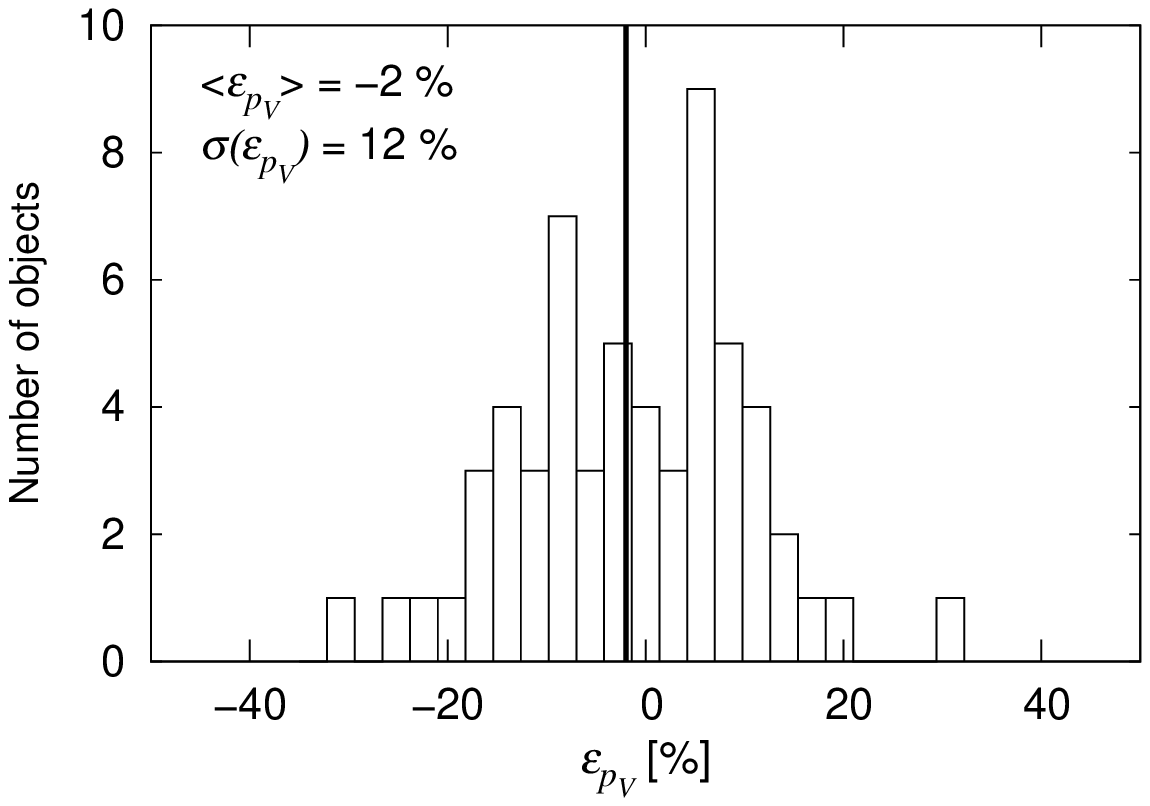}
    \includegraphics[width=68mm]{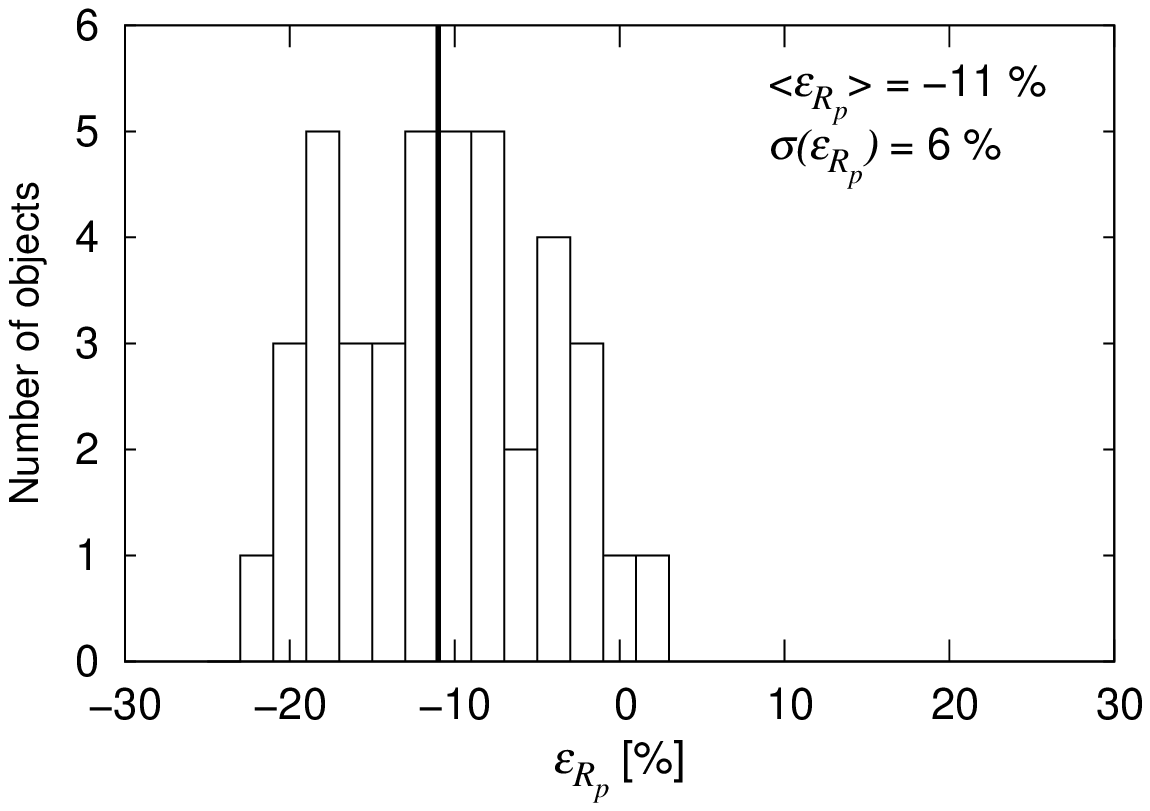}
  \end{center}
  \caption{
    Fractional difference histograms of $D$, $\eta$, $p_V$ and $R_p$. We define $\varepsilon= 100(x - x_M)/x$, where $x$ is the parameter value in this work and $x_M$ is the correspoding value taken from Table 1 by \citet{Masiero2011}. 
    The vertical lines mark the corresponding average values. 
    Note: only parameters resulting from the same input values of $H$ contribute to these histograms.
  }
  \label{fig:histogramse}
\end{figure*}   

Figure \ref{fig:histograms} shows that our parameter determinations and those of \citet{Masiero2011} are compatible in spite of the slight differences in the data set and the thermal modeling used in this work (refer to Sect. \ref{sec:data} and Appendix \ref{app:model}), from which we do not expect to obtain exactly the same best-fit parameters for each object. 
In order to carry out a detailed comparison between our results and those of \citet{Masiero2011}, we computed the mean fractional difference ($\varepsilon$) and corresponding standard deviations of $D$, $\eta$, $p_V$ and $R_p$. 
Let $\varepsilon= 100(x - x_M)/x$, where $x$ is the parameter value for a given object in this work and $x_M$ is the correspoding value taken from Table 1 by \citet{Masiero2011}. 
The distributions of $\varepsilon$ values are plotted in Fig. \ref{fig:histogramse}. 
These histograms only include parameter determinations that have the same $H$ as input in order to identify possible discrepancies in results not caused by different values of $H$.
We find that our values of $D$ and $\eta$ tend to be slightly greater by 1\% and 3\%, respectively, whereas our $p_V$ values are lower by 2\%, though these deviations are small compared to the errorbars. 
On the other hand, there is a large bias towards lower values of $R_p$ that, while still being within the errorbar, must be addressed.

Most probably, the $R_p$ discrepancy is associated with how the reflected flux $r_\lambda$ is calculated. 
In particular, we take the solar flux at 3.4 $\mu$m ($f_{IR_\odot}$ in Eq. \ref{ec:rlc_rlambda}) from the solar power spectrum at zero air mass of Wehrli\footnote{\scriptsize http://rredc.nrel.gov/solar/spectra/am0/wehrli1985.new.html}, based on that by \citet{Neckel1984}. 
Any differences in input, including solar visible magnitude, taken from tabulated data sources that may cause our $r_\lambda$ to be systematically 10\% greater than that of  \citet{Masiero2011} would explain our higher values of $R_p$.
For instance, taking into account that there is only one optimum value of $r_{\mathrm{s},\lambda}$ to fit a given W1 data set, from Eq. \ref{ec:rlc_appendix} it is clear that a larger $r_\lambda$ will have associated a lower best-fit value of $R_p$. 

The Monte Carlo estimations by the NEOWISE team show that the errorbars associated to the fitting of the data are always small compared to the errors inherent to the thermal model itself. 
The relative errors in diameters derived from the NEATM have been characterised to be $\sim$10\%--15\% \citep{Harris2006}. 
From these facts and the widths of the $\varepsilon$-value distributions of Fig. \ref{fig:histogramse}, we consider it safe to assume a minimum relative error of 10\% in diameter and 20\% in beaming parameter, $p_V$ and $R_p$. 
On the other hand, large uncertainties in the absolute magnitude (sometimes as large as $\sim$ 0.3 magnitudes) will also affect the values of $p_V$, so 20\% is probably an optimistic assumption in some cases. 

Finally, we also evaluate how differences in the values of $H$ result in different values of $p_V$ and $R_p$.
We downloaded the MPC orbital element file as of May 2012 and compared the values of absolute magnitude ($H_U$) to those used by  \citet{Masiero2011}, $H_M$. 
About 50000 $H$-values have been updated between these two works, and $\sim $38000 have been enlarged. 
Figure \ref{fig:HmagDiff} shows a histogram of $\Delta H \equiv H_U - H_M$ for the B-types in this work.
Out of the 52 objects with $\Delta H \ne 0$, as many as 43 of them have $\Delta H > 0$.
Our size determinations agree to within 10\%, therefore larger updated values of $H$ will result in lower values of geometric albedos. 
 
\begin{figure}
  \centering
  \includegraphics[width=68mm]{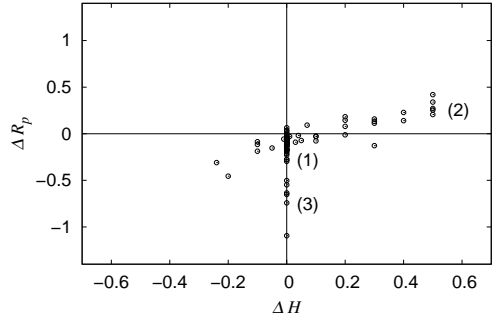}
  \caption{
    Differences in albedo ratio determinations versus difference in absolute magnitude corresponding to the B-types in this paper and those by \citet{Masiero2011}. 
    Note that while the majority of cases verify $\Delta H = 0$, our $R_p$ values tend to be lower (1). 
    On the other hand, some of the differences in albedo ratio (2) are explained by the differences in the updated values of $H$ used in this paper. 
    Finally, some points show very large differences, $\Delta R_p < -0$.5, in the albedo ratio value (3) despite the fact that $\Delta H = 0$ (more details in the text). 
    \label{fig:deltaHdeltaRp}
  }
\end{figure}

In Fig. \ref{fig:deltaHdeltaRp} we show a plot of $\Delta R_p  \equiv R_p - (R_p)_M$ versus $\Delta H$ for all the B-types with determined values of $R_p$.
The notation $(R_p)_M$ refers to the corresponding albedo ratios by \citet{Masiero2011}.
There are three features to note in this plot: 
(1) our values of $R_p$ tend to be $\sim$10\% systematically lower, as we already noted (see Fig. \ref{fig:histogramse});
(2) most points off the $\Delta H = 0$ axis show a direct correlation between $\Delta R_p$ and $\Delta H$, as expected from the discussion above;
(3) some points show $\Delta R_p < -0$.5 even though $\Delta H = 0$.
The points of feature (3) are explained by an inconsistency in the $p_V$ values of \citet{Masiero2011} with their corresponding values of $D$ and $H$: they do not verify Eq. \ref{ec:pVHD} and are always lower than the predicted $p_V$. 

To sum up, we have shown that if the input values of $H$ are equal, our model fits are consistent within the model errorbars with those presented in Table 1 of \citet{Masiero2011}. 
The tendency to 10\%-lower values of $R_p$ is likely caused by differences in solar power spectra data taken to estimate the reflected light component at NIR wavelengths (see Eq. \ref{ec:rlc_rlambda}).  
We have also examined how updated input values of $H$ affect the best-fit parameter values and showed how increasing the value of $H$ results in greater values of $R_p$ and vice versa.

\end{document}